\documentclass[%
reprint,
superscriptaddress,
twocolumn,
notitlepage,
nofootinbib,
amsmath,amssymb,
aps,
prx,
]{revtex4-1}

\usepackage{graphicx}
\usepackage{dcolumn}
\usepackage{bm}

\usepackage{amsthm}
\usepackage{amsmath}
\usepackage{amssymb}
\usepackage{graphicx}
\usepackage{url}
\usepackage{float}
\usepackage{bm}
\usepackage{ifthen}
\usepackage[usenames,dvipsnames]{color}
\usepackage{mathrsfs}
\usepackage[colorlinks=true,citecolor=blue,urlcolor=black]{hyperref}
\usepackage{float}
\usepackage{subfigure}
\usepackage{enumitem}
\usepackage{color}
\usepackage{setspace}
\usepackage{braket}

\usepackage{lipsum}

\usepackage{verbatim}

\begin{document}
	
	\newpage
	
	\title{Asymmetric Protocols for Scalable High-Rate Measurement-Device-Independent Quantum Key Distribution Networks}
	
	\author{Wenyuan Wang}
	\thanks{wenyuan.wang@mail.utoronto.ca}
	\affiliation{Centre for Quantum Information and Quantum Control (CQIQC), Dept. of Electrical \& Computer Engineering and Dept. of Physics, University of Toronto, Toronto,  Ontario, M5S 3G4, Canada}
	
	\author{Feihu Xu}
	\thanks{feihuxu@ustc.edu.cn}
	\affiliation{Shanghai Branch, National Laboratory for Physical Sciences at Microscale, University of Science and Technology of China, Shanghai, 201315, China}
	
	\author{Hoi-Kwong Lo}
	\thanks{hklo@comm.utoronto.ca}
	\affiliation{Centre for Quantum Information and Quantum Control (CQIQC), Dept. of Electrical \& Computer Engineering and Dept. of Physics, University of Toronto, Toronto,  Ontario, M5S 3G4, Canada}

	\begin{abstract}
	Measurement-device-independent quantum key distribution (MDI-QKD) can eliminate detector side channels and prevent all attacks on detectors. The future of MDI-QKD is a quantum network that provides service to many users over untrusted relay nodes. In a real quantum network, the losses of various channels are different and users are added and deleted over time. To adapt to these features, we propose a type of protocols that allow users to independently choose their optimal intensity settings to compensate for different channel losses. Such protocol enables a scalable high-rate MDI-QKD network that can easily be applied for channels of different losses and allows users to be dynamically added/deleted at any time without affecting the performance of existing users.
	\end{abstract}
	
	\date{\today}
	\maketitle

\begin{figure}[h]
	\includegraphics[scale=0.28]{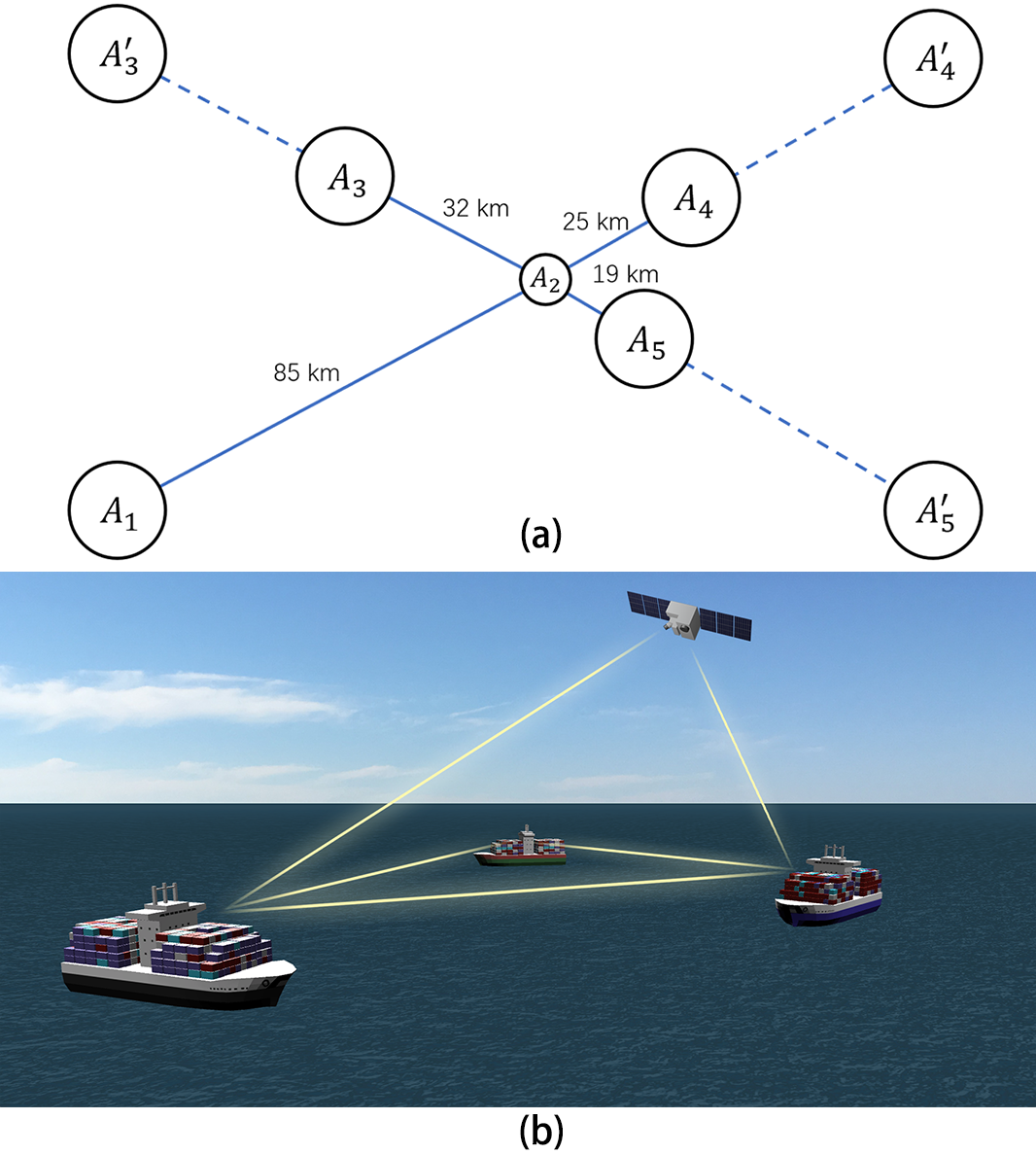}
	\caption{(a) Part of the QKD network setup from Ref.\cite{quantumnetwork1}. Here as an example, we focus on the five nodes with high asymmetry (Nodes $A_1,A_3,A_4,A_5$ connected with $A_2$, corresponding to nodes 1-5 in Ref.\cite{quantumnetwork1}), where $A_2$ can be set up as an untrusted relay. We keep the same topology and redraw it as a star-shaped MDI-QKD network with four users connected to a single untrusted relay. When performing MDI-QKD, all users need to accommodate for the longest channel (i.e. $A_1$) and add losses to their channels (e.g. extending to $A_3', A_4', A_5'$), if previous protocols are used. (b) Ship-to-ship communication and ground-satellite communication, where the participants' distances to the detector are constantly changing, and the channels will thus have quickly varying asymmetry.}
	\label{fig:network}
\end{figure}

\section{Introduction}

Quantum key distribution (QKD) allows two parties to share a pair of random keys with unconditional security. However, while theoretically secure, practical QKD systems have detector side channels, which make them susceptible to attacks from hackers, making detectors the Achilles' Heel of QKD systems~\cite{timeshift,blinding}. The measurement-device-independent (MDI) QKD~\cite{mdiqkd} protocol allows an untrusted third-party to make measurements, thus avoiding all security breaches from detector side channels.

MDI-QKD uses two channels between an untrusted relay Charles and each of Alice and Bob. (Here in this work, we will focus on only discrete-variable MDI-QKD.) Since MDI-QKD depends on two-photon Hong-Ou-Mandel interference, its secure key rate heavily depends on the level of \textit{symmetry} between the two channels, i.e., how close the two channel losses are~\cite{mdipractical,HOM}. Previous experiments of MDI-QKD either were performed in the laboratory over symmetric fibre spools~\cite{mdiexp1,mdiexp2,mdiexperiment,mdi200km,mdiexp3,mdi404km}, or had to deliberately add a tailored length of fibre to the shorter channel (to introduce additional loss) in exchange for better symmetry~\cite{mdiPOP}.

The future of MDI-QKD is to implement a MDI-QKD network - which, like a QKD network, allows many users to securely communicate simultaneously, but does not require trusted relays in the network, which is a huge advantage over traditional point-to-point QKD. Field implementations of point-to-point QKD networks have been reported in e.g., Refs.\cite{quantumnetwork1,quantumnetwork2,quantumnetwork3}. Importantly, all these QKD networks require trusted relays between users that exchange the keys acquired from point-to-point QKD sessions with each user. Notably, Tang et al. \cite{mdinetwork} reported the first (and the only one to date) three-user star-shaped MDI-QKD network experiment in a metropolitan setting. However, the MDI-QKD network experiment also has the limitation of being only feasible for near-symmetric channels. Using the parameter optimization algorithm in~\cite{mdiparameter}, it chooses identical intensities and probabilities for all three users. However, this is only possible because the channels are nearly symmetric, and the key rate for such a protocol will degrade very quickly with an increased level of asymmetry between channels. (Note that, there have also been proposals for continuous variable (CV) MDI-QKD~\cite{CVMDIQKD1,CVMDIQKD2}, which provides high key rate for short distances, but is typically limited to distances $<25km$ even when assuming a high detector efficiency of $98\%$.)

In a realistic setup, a quantum network will very likely have asymmetric channels due to different geographical locations of sites. For instance, the channel losses in Ref.\cite{quantumnetwork1, quantumnetwork2} are largely different. Here we select 5 nodes from the Vienna QKD network~\cite{quantumnetwork1} and show them in Fig.~\ref{fig:network}(a), where the biggest difference between channels is as large as 66km. If we'd like to perform MDI-QKD over these locations, although one can add additional fibres to each channel to compensate for channel differences, users will have to accommodate for the lowest-transmittance channel -- just like in ``\textit{Liebig's barrel}" -- and have sub-optimal rate. Moreover, in a scalable network with large numbers of dynamically added/deleted users, it is not practical to add fibres and maintain symmetry between each pair of users all the time.

Additionally, if one is to implement a MDI-QKD network over free-space between mobile platforms (e.g., satellite-based MDI-QKD~\cite{satelliteQKD} or maritime MDI-QKD between ships), the losses in the channels are constantly changing, and the channels will often be highly asymmetric, as shown in Fig.~\ref{fig:network}(b). 

The issue of MDI-QKD with asymmetric channel losses was first considered in Ref. [4]. which provided a rule of thumb on the ratio of intensities between Alice's and Bob's signals. However, Ref. [4] was restricted to protocols where the intensities of the optical signals are symmetric with respect to two bases, X and Z. In this paper, we make no such assumption.

The key goal of the present paper is designing a software solution that enables high key generation rate in a general scalable MDI-QKD network with arbitrary losses for various channels. {\color{black}More concretely, we propose a type of asymmetric MDI-QKD protocols where intensities are not only different for Alice and Bob, but also different in X and Z bases.} This type of protocols can provide as much as 79 times higher key rate than previous protocols \cite{mdifourintensity} that were designed for symmetric channels. Moreover, it enables a much larger region of possible combinations of channels. For instance, even at a small data size of $N=10^{11}$ ($N$ is defined as the total number of pulses sent by Alice and Bob), one can generate a high secret key rate of $R=10^{-7}$ per pulse even through an extremely asymmetric channel pair of (0km, 90km) for (Alice's, Bob's) channels, whereas with previous protocols no key could be generated at all. This completely removes the requirement of symmetric channels in MDI-QKD.

So far, the optimal decoy state method for MDI-QKD is the 4-intensity protocol proposed by Zhou et al. \cite{mdifourintensity}. In this protocol, Alice and Bob each uses three intensities $\{\mu, \nu, \omega\}$ in the X basis to perform decoy-state analysis \cite{decoystate_LMC,decoystate_Hwang,decoystate_Wang}, and uses one signal intensity $\{s\}$ in the Z basis to generate the secret key. Including the probabilities $P$ for each intensity, the 4-intensity protocol uses the same set of 6 parameters for Alice and Bob\footnote{{\color{black}Although Ref. \cite{mdifourintensity} mentioned on passing the possibility of using different intensities of optical signals for Alice and Bob, little analysis on this important case was performed there. So, up till now, it has not been clear how exactly Alice and Bob could compensate for asymmetric channel losses with different signal intensities.}}:
\begin{equation} 
[s, \mu, \nu, P_s, P_\mu, P_\nu]. \nonumber
\end{equation}

\noindent The 4-intensity protocol can greatly improve MDI-QKD performance under limited data size. However, it limits its discussions to the symmetric case only (optimizing 6 parameters), which is suboptimal in an asymmetric setting. 

In Appendix A we provide an intuitive illustration of why using prior protocols (with same parameters for Alice and Bob) and adding additional fibres to channels are suboptimal when channels are asymmetric, and how we can get better performance by using different intensities for Alice and Bob - which we will describe in detail in the next section.

{\color{black}
Moreover, we outline an important conceptual advance here: a common folklore in the field is that MDI-QKD relies on Hong-Ou-Mandel (HOM) dip and, therefore, it is important to use matched intensities at the beam-splitter of the receiver, Charles, in MDI-QKD. Here, we show that such a folklore is, in fact, a misconception. We consider the \textit{decoupling} of the X-basis from the Z-basis. We note clearly that for MDI-QKD, only the X-basis relies on the indistinguishability of photons from the two beams, while the Z basis does \textit{not} require the indistinguishability of the signals at all. Therefore, if we use the Z-basis to send signals, the quantum bit error rate (QBER) of the signal states are highly insensitive to intensity mismatch at the beam-splitter of the receiver (Charles). This allows us to independently vary the intensities of the users (Alice and Bob) to optimize the key generation rate. The detailed reasons will be described in Subsection II.A.

Overall, the type of protocols we propose have two kinds of inherent asymmetries: the asymmetry between Alice and Bob, and the asymmetry between X and Z bases, which, together, enable the protocol to effectively compensate for different pairs of channels and maintain good key generation rate. We will describe how to optimally choose these asymmetric parameters in Subsection II.B, and present the simulation results to show the effectiveness of our protocol compared with prior protocols in Section III.
}

\section{Asymmetric Protocols}

In this work, we show that it is possible to effectively compensate for channel asymmetry by simultaneously decoupling the X and Z bases and allowing asymmetric intensities for Alice and Bob in decoy-state MDI-QKD. {\color{black}In section II.A, we will provide a theoretical explanation for the advantage of such a type of protocols, and in section II.B we focus on one example implementation, the case where three decoy intensities are used in the X basis (which we denote as ``7-intensity protocol"), and discuss how to perform efficient parameter optimization. Moreover, we also discuss other protocols such as using two-decoys and four-decoys in the X basis to demonstrate the generality of our concept.}

Note that, our method proposed here is a general result that can be applied to any decoy-state MDI-QKD protocol with WCP source, for both asymptotic and finite-size cases. We show in Appendices B and C, that the scaling of key rate versus distance is determined by the signal states, so in principle any number of decoy states can be used so long as they can effectively estimate the single-photon contributions.

\subsection{Concept}

\begin{figure}[!ht]
	\includegraphics[scale=0.18]{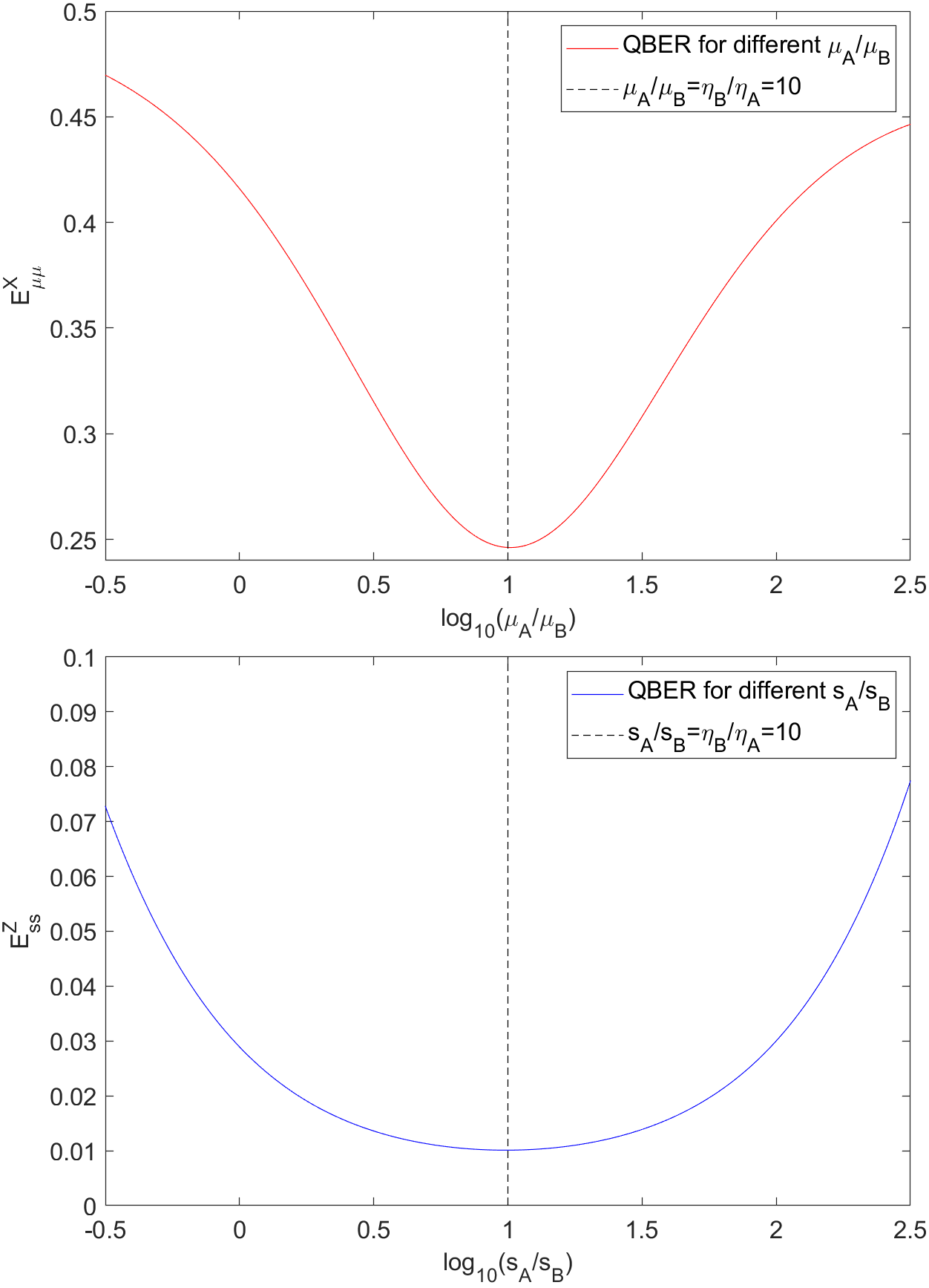}
	\caption{An example of the respective quantum bit error rate (QBER) in X basis and Z basis (i.e. $E^Z_{ss}$ and $E^X_{\mu\mu}$) versus ratio of intensities, for MDI-QKD using WCP sources. Parameters from Table I are used. Here we consider the case where the respective distances from Alice and Bob to Charles are $L_{\text{A}}=60km$, $L_{\text{B}}=10km$ (i.e. the ratio of transmittances in the two channels satisfies $\eta_{\text{B}}/\eta_{\text{A}}=10$). We fix $s_{\text{B}}=0.2$ (or $\mu_{\text{B}}=0.2$) and scan over different $s_{\text{A}}$ (or $\mu_{\text{A}}$). Specifically, we also mark out the position where $s_{\text{A}}\eta_{\text{A}}=s_{\text{B}}\eta_{\text{B}}$ ($\mu_{\text{A}}\eta_{\text{A}}=\mu_{\text{B}}\eta_{\text{B}}$). Because QBER in the X basis heavily depends on the visibility of two-photon interference, it is lowest when intensities arriving at Charles' beam-splitter are equal (Similar observation has been made in Ref.\cite{HOM}.) However, importantly, the Z basis does not require signal indistinguishability, and its QBER is determined mainly by misalignment. The misalignment makes the Z basis QBER also slightly dependent on the interference visibility, and lowest when arriving intensities are equal, but such QBER is much less sensitive to unbalanced intensities and is relatively low even if $s_{\text{A}}\eta_{\text{A}} \neq s_{\text{B}}\eta_{\text{B}}$. Therefore, by decoupling X and Z basis, we can maintain highly balanced decoy state intensities arriving at Charles in the X basis, while further optimizing signal intensities to obtain higher key rate. {\color{black}As a quantitative example of such difference in sensitivity, let us consider $L_{\text{A}}=60km$, $L_{\text{B}}=10km$ and $N=10^{11}$ (Table IV line 1). An optimal key rate of $R=3.1\times 10^{-5}$ can be achieved, where optimal decoy state intensities satisfy ${\mu_A / \mu_B}={\nu_A / \nu_B}=9\approx{\eta_B / \eta_A}$, and $E^X_{\mu\mu},E^X_{\nu\nu}$ are both close to 25\% (see Table IV for the full list of intensities and probabilities). Even a relatively small deviation, such as choosing ${\mu_A / \mu_B}={\nu_A / \nu_B}=10^{0.5}=3.16$ when fixing $\mu_B,\nu_B$ (which results in $E^X_{\mu\mu}$ and $E^X_{\nu\nu}$ close to 32\%), results in zero rate. On the other hand, the optimal signal states satisfy $s_A / s_B=3.5$, which deviates from $\eta_B / \eta_A$, but $E^Z_{\mu\mu}$ is still a rather small $0.013$. In fact, here even if we choose $s_A=s_B=0.2$, we can still get $R=1.0\times 10^{-5}$ while $E^Z_{\mu\mu}=0.029$.}}
	\label{fig:QBER}
\end{figure}

Here, let us first outline the key physical intuition behind how to make a MDI-QKD protocol work effectively when channels are asymmetric.

Consider the key rate formula of MDI-QKD \cite{mdiqkd,mdifourintensity}:
\begin{equation}
\begin{aligned}
R=P_{s_{\text{A}}}P_{s_{\text{B}}} \{(s_{\text{A}} e^{-s_{\text{A}}})(s_{\text{B}} e^{-s_{\text{B}}}) Y_{11}^{X,L}[1-h_2(e_{11}^{X,U})]\\
-f_eQ_{ss}^Z h_2(E_{ss}^Z)\}
\end{aligned}
\end{equation}

\noindent where $Q_{ss}^Z, E_{ss}^Z$ are the gain and QBER in the Z (signal) basis, $Y_{11}^{X,L},e_{11}^{X,U}$ are the lower (upper) bounds of single-photon yield and QBER, estimated from the decoy state statistics in the X basis (i.e. the observed gain and QBER for decoy states $Q_{ij}^X,E_{ij}^X$, where $i,j$ are decoy intensities, such as in $\{\mu_{\text{A}},\nu_{\text{A}},\omega\}$ and $\{\mu_{\text{B}},\nu_{\text{B}},\omega\}$ if Alice and Bob each chooses three decoy states), $h_2$ is the binary entropy function, and $f_e$ is the error-correction efficiency. 

In the key rate formula, the first part corresponds to key generation (where the privacy amplification depends on the single-photon contributions estimated from decoy-state analysis), and the second part corresponds to error-correction for the signal states. 

Here, we make two key observations:\\

{\color{black}(1) For MDI-QKD, only the diagonal (X) basis requires the indistinguishability of the signals from Alice and Bob, while the rectilinear (Z) basis does not.}

(2) In our protocol, the intensities of the signal states $\{s_{\text{A}}, s_{\text{B}}\}$ used in the Z basis are independent from those of the decoy states used in the X basis, which means that the privacy amplification process (to bound Eve's information on the final key, i.e. estimate the phase error rate) in the X basis is completely \textit{decoupled} from error-correction in the Z basis for key generation.\\

{\color{black}Point (1) is because, in MDI-QKD, Charles performs a Bell-state measurement with \textit{post-selection}, making the protocol different from a simple two-photon interference in standard Hong-Ou-Mandel (HOM) dip. Here let us follow the discussions in Ref. \cite{mdiqkd} (and consider the experimental setup from Fig. 1 in Ref. \cite{mdiqkd}). Note that while Alice and Bob randomly send signals in the X and Z bases, Charles always measures in the Z basis (as defined by his polarizing beam-splitter(PBS)) and post-selects detector click events that correspond to the two Bell states  $\ket{\psi^+}=1/\sqrt{2}(\ket{HV}+\ket{VH})$ and $\ket{\psi^-}=1/\sqrt{2}(\ket{HV}-\ket{VH})$. Such a post-selection results in an asymmetry between the two bases. In the Z basis, only events where Alice and Bob sent opposite states (e.g. $\ket{HV}$ or $\ket{VH}$) are accepted as bits. In these cases no photon interference takes place and indistinguishability between the two input photon beams is not required, because each of the clicking detectors respectively only receives signal from either Alice or Bob but never both. For WCP sources, in the ideal case with no misalignment or dark counts, the intensities of the pulses and even their spectrum and timing need not be matching at all. In the X basis, however, the events may correspond to identical states sent by Alice and Bob (e.g. $\ket{++}$ and $\ket{--}$ corresponding to $\ket{\psi^+}=1/\sqrt{2}(\ket{++}-\ket{--})$), which do interfere at the beam splitter.\footnote{{\color{black}Another case where Alice and Bob sent $\ket{+-}$ or $\ket{-+}$ corresponds to the other Bell state, $\ket{\psi^-}$. A two-photon interference happens not at the beam splitter but at the polarizing beam splitter (PBS) instead. This setup is slightly different from HOM interference but similar to that of Ref. \cite{Photon_Interference}, and also requires indistinguishability of e.g. spectrum, timing, and matching intensities. For simplicity, here we will use the term ``two-photon interference" for both cases.}} To ensure that the correct events are triggered, a good visibility of such a two-photon interference is required. Note that for WCP sources, the interference visibility is at most 50\% (resulting in a 25\% observed QBER for $E^X_{\mu\mu},E^X_{\nu\nu}$ even in the ideal case, but we can perform decoy-state analysis to correctly estimate a low $e^{X,U}_{11}$ for the single photon components) and the visibility will quickly drop when intensities are mismatched, such as observed in \cite{HOM}. 

Therefore, a low QBER in the X basis heavily relies on the indistinguishability of the signals and the balance of incoming intensities at Charles, while such dependence is not present in the Z basis.\footnote{{\color{black}In the non-ideal case with basis misalignment, there may be a slight dependence in the Z basis too, as we see in Fig. \ref{fig:QBER}, because misalignment results in crosstalk between signals from the two bases, but it will be a much smaller dependence than that in the X basis.}} Such a conclusion is rather general and also not dependent on the degree-of-freedom used for qubit encoding - such as polarization encoding or time-bin phase encoding (where $\ket{HV}$ and $\ket{VH}$ in the Z basis correspond to pairs of early and late pulses, which will similarly not interfere at the beam splitter since they have different timing).}

Now, having explained the reason behind point (1), let us discuss how the parameter choices in point (2) impact the MDI-QKD protocol. For the decoy states, their role is to estimate the single-photon contributions as accurately as possible. As mentioned above, when channels are asymmetric, using same intensities for Alice and Bob (hence different intensities arriving at Charles after the channels' attenuation) will result in poor interference visibility and high QBER in the X basis, and consequently poor estimation of $e^{X,U}_{11}$. For a good interference visibility, Alice and Bob should try to maintain similar intensities arriving at Charles, so the decoy intensities should be chosen to roughly satisfy
\begin{equation}
	\mu_{\text{A}}\eta_{\text{A}}=\mu_{\text{B}}\eta_{\text{B}}
\end{equation}

\noindent where $\eta_{\text{A}}$ and $\eta_{\text{B}}$ are the channel transmittances in Alice's and Bob's channels. A similar equation holds true for $\nu_{\text{A}}$ and $\nu_{\text{B}}$.

For the signal states, they are not involved in privacy amplification. On the other hand, they affect the signal state gain and QBER $Q^Z_{ss}, E^Z_{ss}$ (which determine the amount of error-correction), and the probability of sending single photons for key generation $s_{\text{A}} e^{-s_{\text{A}}} s_{\text{B}} e^{-s_{\text{B}}}$. The key point is, the QBER $E^Z_{ss}$ does \textit{not} require indistinguishability of the signals. If there is no misalignment or noise, $E^Z_{ss}$ would be zero regardless of incoming intensities. In practice, due to imperfections such as misalignment, the QBER $E_{ss}^Z$ (whose full expression can be found in Appendix C Eq. C3) still slightly depends on channel asymmetry and is also minimal if incoming intensities at Charles are balanced - but this is for a much different reason (due to misalignment) than that in the X basis (mostly due to two-photon interference). Furthermore, $E_{ss}^Z$ is much less sensitive to channel asymmetry than QBER in the X basis. We can observe this from Fig.\ref{fig:QBER}.

Note that, not only do signal intensities affect the signal state QBER, they also determine the probabilities of sending single photons, hence affecting key generation too. This means that, while having similar received signal intensities at Charles is surely one important criterion in achieving good key rate, the optimal choice of signal state intensities requires a trade-off between the single photon probabilities and the error correction (and their optimal values can be found by numerical optimization). Generally speaking, the ratio of signal intensities ${s_{\text{A}} / s_{\text{B}}}$ does not satisfy a similar relation as Eq. (2), i.e. generally

\begin{equation}
s_{\text{A}}\eta_{\text{A}}\neq s_{\text{B}}\eta_{\text{B}}
\end{equation}

Therefore, the protocols we propose have two inherent asymmetries: an asymmetry between Alice and Bob (so that they can have different intensities, and establish good two-photon interference in the X basis), and an asymmetry between the X and Z bases (which allows decoy and signal states to be independently optimized). Such inherent asymmetries in the protocols allow us to have novel choice of parameters and maintain good key rate of MDI-QKD, even when Alice's and Bob's channels have very different levels of loss. A more detailed discussion on how such independent choices of decoy and signal states affect the key rate can be found in Appendix D.

{\color{black}Note that, the security of such a protocol with decoupled bases and asymmetric intensities is also not compromised compared to prior art protocols. We make two key assumptions: (1) Given the same photon number n in a pulse, Eve has no way of differentiating the decoy states from signal states in the same basis, and (2) the single photons pairs in X and Z bases cannot be distinguished from each other. The first assumption ensures the decoy-state analysis works even with asymmetric intensities, and the second assumption ensures that decoupling of bases works. More details can also be found in Appendix D.}

In the next subsection we will discuss how to actually choose the optimal decoy and signal intensities, and introduce the main challenge in implementing such asymmetric protocols - performing efficient parameter optimization over a huge parameter space - and how we address this problem by proposing two important theoretical results for the key rate function of asymmetric MDI-QKD, and using them to design an efficient optimization algorithm.

\subsection{Parameter Optimization}

{\color{black}The results in the previous subsection are general and not limited to the number of decoys Alice and Bob use in the X basis. For instance, while using signal states $\{s_A,s_B\}$ in the Z basis, in the X basis Alice and Bob can each use a different set of two decoy states $\{\mu,\nu\}$, three decoy states $\{\mu,\nu,\omega\}$, or even four decoy states $\{\mu,\nu,\nu_2,\omega\}$. The concept of asymmetric intensities between Alice and Bob can in principle also be applied to prior art protocols with non-decoupled bases, such as in Refs. \cite{mdipractical,decoyMDI_Wang} (where Alice and Bob use the same three decoy states $\{\mu,\nu,\omega\}$ for both bases, and the Z basis $\mu$ is used as the signal state for key generation) - it is just that such protocol will have lower key rate since $\mu$ cannot simultaneously satisfy asymmetry compensation and key rate optimization. 
	
As an example, in Table II we list a comparison between the key rate of using different number of decoy states (with and without asymmetric intensities between Alice and Bob) in the presence of asymmetric channels. We include the non-decoupled-bases case \cite{mdipractical,decoyMDI_Wang} too. We can see that, regardless of the protocol, using asymmetric intensities between Alice and Bob always provides higher key rate when channels are asymmetric. Also, the three-decoy case provides significant performance improvement over either the two-decoy case or prior art protocol (which also has three decoy states, meaning that decoupled bases is crucial in the compensation for channel asymmetry). While the asymmetric four-decoy case can provide highest key rate, it provides a limited performance increase (60\%) over three-decoy, but comes at a cost of more complex experimental implementation as well as more difficult data collection and analysis. See Appendix E for a more detailed comparison between the protocols. Overall, we can see that the three-decoy case provides a good balance between ease of implementation and performance. 

Therefore, for practicality here, in the following text we will focus on the three-decoy case as a concrete example (whose symmetric case is the 4-intensity protocol), and generalize it to the asymmetric case by allowing Alice and Bob to have independent intensities and probabilities. This enables a ``7-intensity protocol" (with 3 independent $\{s,\mu,\nu\}$ for each of Alice and Bob, and the vacuum state $\omega_{\text{A}}=\omega_{\text{B}}=\omega=0$) in the asymmetric case.}

For such a protocol, efficient and accurate parameter optimization is crucial for obtaining good key rate (especially when considering the finite-size effects). For the 7-intensity protocol we need to use a total of 12 parameters for a full finite-size parameter optimization:

\begin{equation} 
\vec{v}=[s_{\text{A}}, \mu_{\text{A}}, \nu_{\text{A}}, P_{s_{\text{A}}}, P_{\mu_{\text{A}}}, P_{\nu_{\text{A}}}, s_{\text{B}}, \mu_{\text{B}}, \nu_{\text{B}}, P_{s_{\text{B}}}, P_{\mu_{\text{B}}}, P_{\nu_{\text{B}}}]. \nonumber
\end{equation}

\noindent here we denote the parameters as a vector $\vec{v}$, and when all devices and channel parameters (e.g. channel loss, misalignment, dark count rate, detector efficiency, etc.) are fixed, the key rate is a function of the intensities and probabilities $R(\vec{v})$, and the question of intensity parameter optimization can be viewed as searching for:

\begin{equation}
\vec{v}_{\text{opt}}=\text{arg max}_{\vec{v} \in V}[R(\vec{v})]
\end{equation}

\noindent where $V$ is the search space for the parameters.

To provide a high key rate under finite-size effects, optimal choice of parameters is very important in implementing the protocol. However, the 7-intensity protocol has an extremely large parameter space of 12 dimensions, for which a brute-force search is next to impossible. Therefore, to efficiently search over the parameters in reasonable time, a local search algorithm must be applied. But, as we will show here, an important characteristic of asymmetric MDI-QKD is the discontinuity of first-order derivatives for the function $R(\vec{v})$ with respect to the intensity parameters in $\vec{v}$. This means that a straightforward local-search algorithm, such as previously proposed in~\cite{mdiparameter}, will inevitably fail to find the optimal point, since it requires continuous first-order derivatives of the searched function.

Here we will present two important theoretical results for the key rate versus parameter function, and propose a method to circumvent the problem of discontinuous derivatives and perform efficient and correct local search in parameter space. This method helps us overcome the biggest challenge in successfully implementing the 7-intensity protocol.

\begin{figure}[h]
	\includegraphics[scale=0.35]{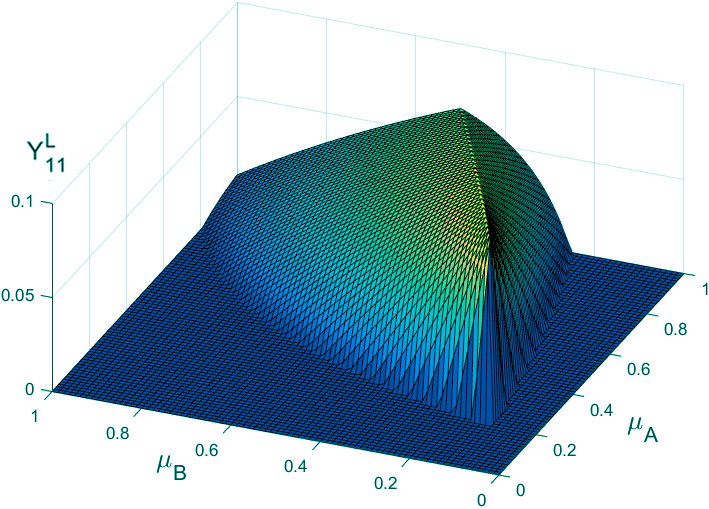}
	\caption{An example of the discontinuity of first-order derivatives of $Y_{11}^{L}$ vs $\mu_{\text{A}}, \mu_{\text{B}}$ function in decoy-state MDI-QKD, for fixed values of $\nu_{\text{A}}=0.2$, $\nu_{\text{B}}=0.1$. Note the ridge on the line ${\mu_{\text{A}} \over \mu_{\text{B}}}={\nu_{\text{A}} \over \nu_{\text{B}}}=2$.}
	\label{fig:ridge}
\end{figure}

Firstly, we propose that there is an inherent symmetry constraint for the \textit{ratio} of optimal decoy intensities, that\\

\textbf{Theorem I}. \textit{for any arbitrary choice of device and channel parameters, the optimal decoy intensities $\mu_{\text{A}}^{\text{opt}}, \nu_{\text{A}}^{\text{opt}}, \mu_{\text{B}}^{\text{opt}}, \nu_{\text{B}}^{\text{opt}}$  that maximize the key rate always satisfy the constraint:}

\begin{equation}
{\mu_{\text{A}}^{\text{opt}} \over \mu_{\text{B}}^{\text{opt}} } = { \nu_{\text{A}}^{\text{opt}} \over \nu_{\text{B}}^{\text{opt}}}
\end{equation}

Secondly, we make an important observation that,\\

\textbf{Theorem II}. \textit{The key rate versus ($\mu_{\text{A}}, \mu_{\text{B}}$) function, for any given $\nu_{\text{A}}, \nu_{\text{B}}$, does not have continuous first-order derivatives.\\}

Both of these theorems result from the fact that the lower bound for single-photon yield, $Y_{11}^{L}$, in decoy-state analysis (whose expression can be found in Ref. \cite{mdipractical,decoyMDI_Wang}) is a piecewise function that depends on whether ${\mu_{\text{A}} \over \mu_{\text{B}}} \leq {\nu_{\text{A}} \over \nu_{\text{B}}}$, where a boundary line ${\mu_{\text{A}} \over \mu_{\text{B}}} = {\nu_{\text{A}} \over \nu_{\text{B}}}$ exists.

Theorem I states that, the optimal parameters that maximize the key rate must lie exactly on this boundary line, while Theorem II states that, the key rate does not have a continuous partial derivative with respect to $\mu_{\text{A}}$ or $\mu_{\text{B}}$ across this boundary line. This will cause the boundary line to behave like a sharp ``ridge", on which the gradient is not defined. An illustration for this ``ridge" can be seen in Fig. \ref{fig:ridge}. A rigorous proof for Theorems I and II in the asymptotic limit can be found in Appendix F.

\begin{figure}[b]
	\includegraphics[scale=0.195]{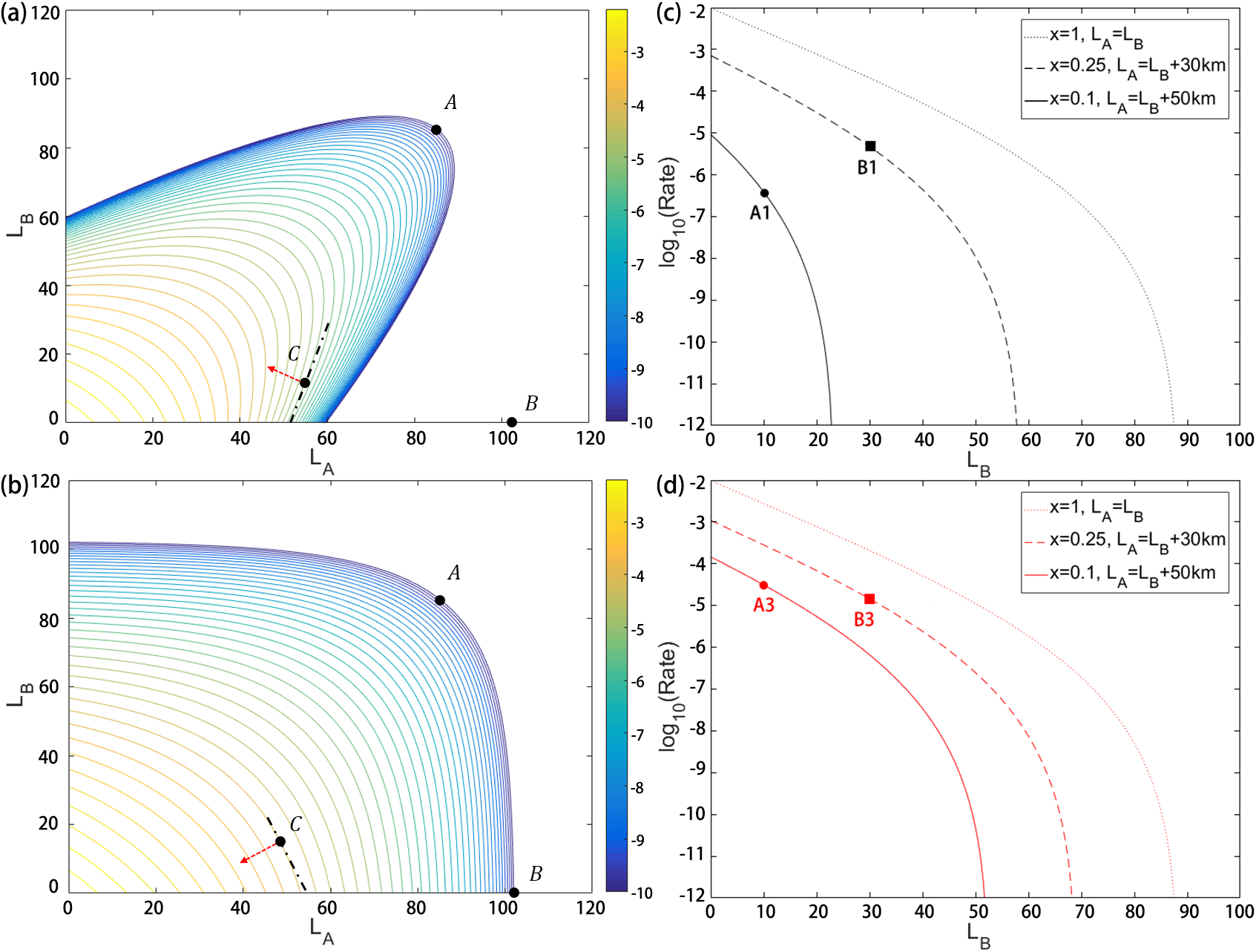}
	\caption{Left: Comparison of rate vs $(L_{\text{A}}, L_{\text{B}})$. The rates are plotted in contours in log-scale, from $10^{-2}$ to $10^{-10}$. We use the parameters from Table I, and $N=10^{11}$. (a) using a previous 4-intensity protocol, (b) using our 7-intensity protocol. As can be seen, while 4-intensity MDI-QKD is limited to only high-symmetry regions, using 7-intensity can greatly increase the applicable region of MDI-QKD, even in extremely asymmetric regions such as $(L_{\text{A}},0)$ where one channel has zero distance (point B). Moreover, we see that with 7-intensity protocol both $L_{\text{A}}, L_{\text{B}}$ components of the gradient for key rate (red dotted arrow) are always negative, meaning that with 7-intensity protocol it is always optimal to only adjust the intensities, and never necessary to add any fibre, while for 4-intensity protocol, adding fibre (e.g. increasing $L_{\text{B}}$ at point C) will sometimes increase the rate. Right: Comparison of rate vs distance (Bob to Charles) for various fixed levels of mismatch $x={\eta_\text{A}\over\eta_\text{B}}$ where $\eta_\text{A},\eta_\text{B}$ are the channel transmittances, (c) using 4-intensity protocol (d) using 7-intensity protocol. As can be observed, the higher the mismatch, the more advantage 7-intensity protocol has (and only when the channels are symmetric will the two protocols perform identically). Data points from A1, A3, B1, B3 from Table III are also shown in the plots.}
	\label{fig:2d_Results}
\end{figure}

{\color{black}
Using Theorems I and II it is possible to transform the coordinates of the search variables, and eliminate the undefined gradient problem of the key rate function. More specifically, instead of expressing $(\mu_{\text{A}},\mu_{\text{B}}),(\nu_{\text{A}},\nu_{\text{B}})$ in Cartesian coordinates, we can express them in polar coordinates $(r_\mu^{\text{polar}}, \theta_\mu^{\text{polar}}), (r_\nu^{\text{polar}}, \theta_\nu^{\text{polar}})$, where polar angles satisfy $\theta_\mu^{\text{polar}}=\theta_\nu^{\text{polar}}$ due to Theorem I. This means we can \textit{jointly} search for:
\begin{equation}
	\theta_{\mu\nu}^{\text{polar}}=\theta_\mu^{\text{polar}}=\theta_\nu^{\text{polar}}
\end{equation} 
\noindent with respect to which the key rate is a smooth function (graphically, this is because we are now always searching along the "ridge").} In Appendix G, we will describe in more detail how to perform efficiently an optimization of the parameters based on local-search to obtain a high secure key rate for our 7-intensity protocol. Our method allows extremely fast and highly accurate optimization for asymmetric MDI-QKD, and takes below 0.1s for each full local search (at any given distance) on a quad-core i7-4790k@4.0GHz PC. Such computing efficiency makes it possible for real-time optimization of intensities on-the-field, and also makes possible a dynamic MDI-QKD network that might add/delete new user nodes in real time. {\color{black}In addition, in Appendix G we also discuss the effect of inaccuracies and fluctuations of the intensities and probabilities on the key rate, and show that our method is robust even in the presence of inaccuracies and fluctuations of the parameters.}

In summary, using our two Theorems and switching to polar coordinates as in Eq. (6) allow us to greatly simplify the optimization problem and allow standard coordinate descent method to be applied here.


\begin{table}[h]
	\caption{Parameters for numerical simulations, adopted from~\cite{mdi404km}, including detector dark count rate and efficiency $Y_0$, $\eta_d$, optical misalignment $e_d$, error-correction efficiency $f$, and failure probability $\epsilon$.}
	\begin{center}
		\begin{tabular}{ccccc}	
			\hline	\hline
			$Y_0$ & $\eta_d$ & $e_d$ & $f$ & $\epsilon$\\
			\hline
			$8 \times 10^{-7}$ & 65\% & 0.5\% & 1.16 & $10^{-7}$\\
			\hline \hline
		\end{tabular}
	\end{center}
\end{table}

\begin{table*}[t]
{\color{black}
	\caption{Example key rate comparison among MDI-QKD protocols where Alice and Bob use different numbers of decoy states in X basis (and each keeps one signal state in the Z basis). The protocol in Ref. \cite{mdipractical,decoyMDI_Wang} where bases are not decoupled is also included for comparison. We use parameters from Table I, $L_{\text{A}}=60km$, $L_{\text{B}}=10km$, and $N=10^{11}$. We can see that, regardless of the protocol, using asymmetric intensities between Alice and Bob always provides higher key rate when channels are asymmetric. The three-decoy protocol has significantly higher key rate than either the prior art protocol (which also uses three decoy states but uses non-decoupled bases) or two-decoy case. While the asymmetric four-decoy case can provide highest key rate, it provides a limited performance increase of 60\%, but comes at a cost of a more complex experimental implementation and more difficult data collection and analysis. Therefore, in the presence of channel asymmetry, the three-decoy case, whose asymmetric case corresponds to the ``7-intensity protocol" (marked in bold), provides a good trade-off between ease of implementation and performance.} 
	\begin{center}
		\begin{tabular}{ccccccc}		
			\hline \hline
			Parameters & prior art protocol in \cite{mdipractical,decoyMDI_Wang} & two-decoy & three-decoy & four-decoy \\
			\hline
			Symmetric & $6.834 \times 10^{-10}$& $0$ & $3.890\times 10^{-7}$ & $1.057 \times 10^{-5}$\\
			Asymmetric & $5.378\times 10^{-7}$ & $7.715 \times 10^{-6}$ & $\boldsymbol{3.106 \times 10^{-5}}$ & $4.932 \times 10^{-5}$\\
			\hline \hline
		\end{tabular}
	\end{center}
}
\end{table*}

\begin{table*}[t]
	\caption{Simulation results for asymmetric MDI-QKD in two scenarios: case A (10km, 60km) and case B (30km, 60km), using parameters from Table I and $N=10^{11}$. {\color{black}We define channel mismatch as $x={{\eta_{\text{A}}}\over{\eta_{\text{B}}}}$ where $\eta_{\text{A}},\eta_{\text{B}}$ are the channel transmittances. Note that in reality, Alice and Bob cannot modify the physical channels, and they can either add loss to the channels or keep them as-is, but cannot decrease channel loss.} Three strategies are compared here: A1 and B1 represent using the old 4-intensity protocol directly. A2 and B2 (not in Fig. \ref{fig:2d_Results}) represent adding fibre to the shorter channel to match the longer channel, i.e. making the channels (60km, 60km). And A3, B3 represent using our new 7-intensity protocol without modifying the channels. As shown here, 7-intensity protocol always returns higher rate than both strategies using 4-intensity protocol.}
	\begin{center}
		\begin{tabular}{ccccccc}		
			\hline \hline
			Protocol & Point &$x$ & $L_{\text{B}}$ & $L_{\text{A}}$ & Rate & Comparison with 4-intensity protocol\\
			\hline
			4-intensity protocol & A1 & 0.1 & 10km & 60km & $3.891 \times 10^{-7}$ & -  \\
			4-intensity protocol + fibre & A2 & 1 & 60km & 60km & $1.862 \times 10^{-6}$ & +379\% \\
			Our protocol  & A3 & 0.1 & 10km & 60km & $3.106 \times 10^{-5}$ & +7883\% \\
			
			4-intensity protocol & B1 & 0.25 & 30km & 60km &  $4.746 \times 10^{-6}$ & - \\
			4-intensity protocol + fibre & B2 & 1 & 60km & 60km & $1.862 \times 10^{-6}$ & -61\% \\
			Our protocol & B3 & 0.25 & 30km & 60km &  $1.445 \times 10^{-5}$ & +204\% \\
			\hline \hline
		\end{tabular}
	\end{center}
\end{table*}

\section{Simulation Results}

{\color{black}
Now, we can proceed to study the performance of asymmetric MDI-QKD protocols with full parameter optimization. Again, we use the 7-intensity protocol as a concrete example as it provides a good trade-off between performance and practicality. We also include simulation results for protocols with alternative numbers of decoy states in Appendix E. 
}

In the main text we focus on the practical case of having finite data size. The asymptotic case of infinitely many data size (and an analytical understanding of the ideal infinite-decoy case) is discussed in Appendices B and C, and its simulation results can be found in Fig.\ref{fig:contours_compare}. 

Our finite-key analysis is described in more detail in Appendix H. For simplicity we consider a standard error analysis in numerical simulations, but it is important to note that our theory is fully compatible with composable security. See Appendix H for discussions. {\color{black}In addition, compared to the "joint-bound" analysis as proposed in Ref. [21] (which jointly considers the statistical fluctuation of multiple observables. Such an analysis model increases the key rate, but introduces multiple maxima undesirable for local search), in the main text here we have chosen to use an "independent-bound" analysis for our simulations, which considers each variables' statistical fluctuations independently, and is far more stable and faster in simulations. However, we specifically note here that all our methods are fully compatible with joint-bound analysis. We list some representative results generated with joint-bound analysis in Table IV for comparison, and will discuss the different finite-size analysis models in more detail in Appendix H.}

Firstly, we consider the key rate for an arbitrary combination of $(L_\text{A}, L_\text{B})$, and perform a simulation of key rate over \textit{all} possible range of Alice and Bob's channels. This provides a bird's-eye view of how using 7-intensities can affect the performance in asymmetric channels. We show the results in Fig.~\ref{fig:2d_Results} (a)(b). From the plot we can make three important observations:

(1) Using 7-intensity protocol, we have a much wider applicable region for asymmetric MDI-QKD, and acceptable key rate can be acquired even for highly asymmetric channels. In addition, 7-intensity protocol will always provide higher key rate than 4-intensity protocol, except when channels are already symmetric.

(2) No matter what position one is at, there is never any necessity for adding loss when 7-intensity protocol is used, and optimizing on-the-spot always provides highest rate. Details can be seen in Fig.~\ref{fig:2d_Results} caption.

(3) Using 7-intensity protocol, even extremely asymmetric scenarios, such as $(L, 0)$ where $L_{\text{B}}=0$, can be used to generate a good key rate. In fact, this provides an even higher rate than with symmetric channels such as $(L, L)$ (As the comparison between points A and B in Fig.~\ref{fig:2d_Results}).

Point 3 has an important practical implication: it can lead to a new type of ``single-arm" MDI-QKD setup. More details can be found in Appendix I and Fig. \ref{fig:singlearm}.

{\color{black}
Here for Points 1-3, we have a good physical understanding of why allowing different intensities for Alice and Bob can provide a larger region where key rate is positive. As discussed in Section II.A, MDI-QKD requires highly balanced intensities arriving at Charles on the two arms in the X basis for good HOM interference, as well as roughly similar (but not necessarily balanced) levels of arriving intensities in the Z basis, which optimize a trade-off between error-correction and probability of sending single-photons. (The optimal choice of intensities is subject to numerical optimization as described in Subsection II.B). Prior methods with same intensities for Alice and Bob will suffer from high QBER both in X and Z basis, while our method decouples X and Z basis, and optimally chooses Alice and Bob's signal and decoy intensities respectively to compensate for channel asymmetry in both bases, ensuing low QBER and allowing for much higher key rate under channel asymmetry. Such effect is present in both asymptotic and finite-key scenarios, and is the underlying reason that the 7-intensity protocol can allow high-rate MDI-QKD regardless of channel asymmetry.}

\begin{table*}[t]
	\caption{{\color{black}Examples of optimal parameters for the 7-intensity protocol, using simulation parameters from Table I. The numerical values are rounded to the accuracy of 0.001 in the table here. As can be observed, Alice and Bob's intensities compensate for channel asymmetry, while their intensity probabilities are mostly identical - since the intensities have already compensated for the asymmetry - despite having have some numerical noises (as the key rate is not sensitive to the probabilities near the maximum, the algorithm satisfies with them having close enough, rather than perfectly identical, values, so the optimal values found are still slightly different even when $x=1$). As shown in Section II, the optimal decoy state ratios are the same, i.e. ${\mu_{\text{A}}\over\mu_{\text{B}}}={\nu_{\text{A}}\over\nu_{\text{B}}}$. Moreover, we can observe that the ratio of decoy states more closely follows $1\over x$ than the ratio of signal intensities.}}
	\begin{center}
		{\color{black}
		\begin{tabular}{cccccccccccccccc}			
			\hline \hline
			$L_{\text{A}}$ & $L_{\text{B}}$ & x & $s_{\text{A}}$ & $\mu_{\text{A}}$ & $\nu_{\text{A}}$ & $P_{s_{\text{A}}}$ & $P_{\mu_{\text{A}}}$ & $P_{\nu_{\text{A}}}$ & $s_{\text{B}}$ & $\mu_{\text{B}}$ & $\nu_{\text{B}}$ & $P_{s_{\text{B}}}$ & $P_{\mu_{\text{B}}}$ & $P_{\nu_{\text{B}}}$ & $R$\\
			\hline
			60km & 10km & 0.1 & 0.662 & 0.522 & 0.100 & 0.600 & 0.033 & 0.255 & 0.202 & 0.058 & 0.011 & 0.600 & 0.031 & 0.256 & $3.106 \times 10^{-5}$  \\
			60km & 30km & 0.25 & 0.593 & 0.457 & 0.089 & 0.581 & 0.036 & 0.266 & 0.294 & 0.125 & 0.024 & 0.580 & 0.034 & 0.269 & $1.445 \times 10^{-5}$  \\
			60km & 60km & 1 & 0.402 & 0.305 & 0.063 & 0.478 & 0.047 & 0.330 & 0.402 & 0.305 & 0.063 & 0.480 & 0.047 & 0.329  & $1.862 \times 10^{-6}$  \\
			\hline \hline
			
		\end{tabular}
		}
	\end{center}
\end{table*}

Additionally, we show that, when channels are highly asymmetric, the asymptotic key rate of the 7-intensity protocol scales quadratically with the \textit{lower} transmittance among the two channels - which means that, albeit always being able to provide higher key rate and being much more convenient than e.g. adding fibres when channels are asymmetric (which is a relation we rigorously prove in Appendix C.2 for the asymptotic case), the 7-intensity protocol will not change the asymptotic scaling properties of MDI-QKD key rate, which is still quadratically related to transmittance - {\color{black}physically, this is understandable, since although we effectively compensate for channel asymmetry with optimized intensities and allow good Hong-Ou-Mandel interference at Charles for decoy states, MDI-QKD still fundamentally depends on two single signal photons both passing through the channels, hence its key rate is quadratically related to transmittance, even in the asymmetric case with compensated intensities.} More detailed discussions and analytical proofs of the above observations can be found in Appendices B and C.

Now, as a concrete example, let us consider two sets of channels at $(L_{\text{B}}=10km, L_{\text{A}}=60km)$ and $(L_{\text{B}}=30km, L_{\text{A}}=60km)$, through which Alice and Bob would like to perform MDI-QKD. We compare strategies of using the 4-intensity protocol, directly or with fibres added until channels are symmetric, with directly using our 7-intensity protocol. As can be seen in Table III, using 7-intensity protocol can sometimes provide as much as 79 times higher key rate, and its rate is also always higher than either strategies with 4-intensity protocol.

In fact, we can also show this by plotting key rate vs $L_{\text{B}}$ under a fixed mismatch {\color{black}$x={{\eta_{\text{A}}}\over{\eta_{\text{B}}}}$ where $\eta_\text{A},\eta_\text{B}$ are the channel transmittances} (i.e. fixed difference between $L_{\text{A}}$ and $L_{\text{B}}$). This is also the scenario studied by Ref. \cite{mdipractical}. Results are shown in Fig.~\ref{fig:2d_Results} (c) and (d). The data points A1/A3 and B1/B3 in Table III are also plotted. As can be seen, the higher the asymmetry between channels, the more improvement we can gain from using 7-intensity protocol.

\begin{figure}[h]
	\includegraphics[scale=0.27]{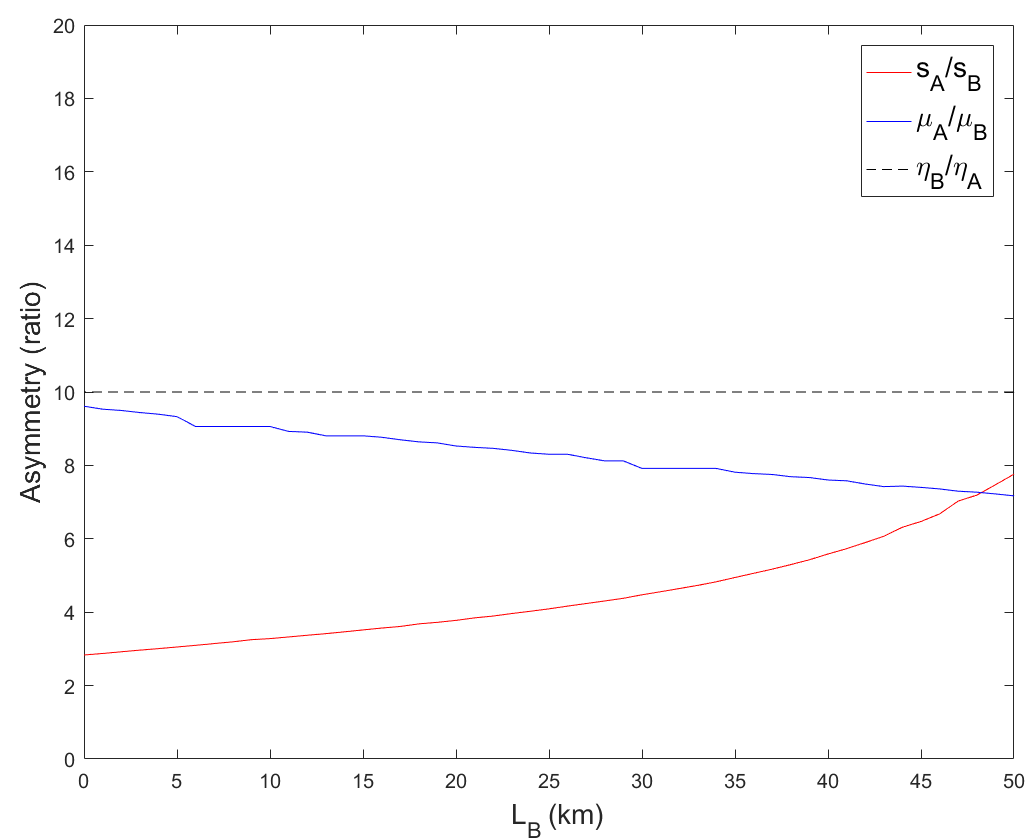}
	\caption{{\color{black}Here we plot the ratios of signal intensities and decoy intensities versus distance, when channel mismatch is fixed at $x=0.1$ (i.e. $L_A=L_B+50km$). The simulation parameters are from Table I (and this plot of intensities corresponds to the solid red key rate line in Fig. \ref{fig:2d_Results} (d)) We also include the line ${\eta_{\text{B}} \over \eta_{\text{A}}} = 10$ for comparison. We can observe that, the ratio of decoy states roughly follows ${\eta_{\text{B}} \over \eta_{\text{A}}}$ (to maintain good HOM interference visibility in X basis), while the optimal ratio of signal intensities varies greatly between $1$ (optimal for probability of sending single photon) and ${\eta_{\text{B}} \over \eta_{\text{A}}}$ (optimal for $E^Z_{ss}$). This is because signal states affect both key generation and error-correction, so having similar intensities arriving at Charles after channel attenuation is not the only criteria for good key rate, and optimal parameters do not necessarily satisfy $s_{\text{A}}\eta_{\text{A}} = s_{\text{B}}\eta_{\text{B}}$. In fact, since signal states in Z basis are decoupled from X basis, and $E^Z_{ss}$ is less sensitive to unbalanced arriving intensities, $s_{\text{A}}\over s_{\text{B}}$ can be much more freely optimized between $1$ and ${\eta_{\text{B}} \over \eta_{\text{A}}}$, allowing 7-intensity protocol to have higher key rate.}}
	\label{fig:ratio}
\end{figure}

{\color{black}
Here, we also list some examples of optimal parameters found by the optimization algorithm, which are listed in Table IV. As we can observe from the table, Alice and Bob adjust their intensities to compensate for channel asymmetry. Physically, since MDI-QKD depends on Hong-Ou-Mandel interference of two WCP sources in the X basis, we expect the \textit{received} intensity for decoy state at Charles to be similar on the two arms to ensure good visibility (and consequently lower QBER) in the X basis, i.e. the ratio of decoy intensities ${\mu_{\text{A}} \over \mu_{\text{B}}}$ and ${\nu_{\text{A}} \over \nu_{\text{B}}}$ would roughly follow the rule-of-thumb of $\mu_{\text{A}}\eta_{\text{A}}=\mu_{\text{B}}\eta_{\text{B}}$, which is indeed what we can observe from Table IV and Fig.\ref{fig:ratio}.
	
On the other hand, the ratio of signal intensities ${s_{\text{A}} \over s_{\text{B}}}$ deviates more from ${\eta_{\text{B}} \over \eta_{\text{A}}}$. This is because, as mentioned in Subsection II.A, signal intensities not only affect the Z basis QBER, but also need to optimize a trade-off between the single photon probabilities and error-correction. This makes it usually not follow $s_{\text{A}}\eta_{\text{A}}=s_{\text{B}}\eta_{\text{B}}$. An illustration of the ratios of decoy intensities and signal intensities can be seen in Fig. \ref{fig:ratio}.
}


Now, having demonstrated the new 7-intensity protocol, we proceed to introduce a powerful reality application for it: a scalable high-performance MDI-QKD network where any node can be dynamically added or deleted. We consider the channels from a real quantum network setup in Vienna, reported in Ref.\cite{quantumnetwork1}. We focus here on the high-asymmetry nodes, $A_1,A_2,A_3,A_4,A_5$, as shown in Fig. \ref{fig:network}(a). We found that our method leads to much higher key rates, and allows easy dynamic addition or deletion of nodes. Since intensities can be independently optimized for each pair of channels, the establishment of new connections does not affect any existing connections, hence providing good scalability for the network (compared to e.g. the case of using 4-intensity protocol with the strategy of adding fibres, where each channel needs to accommodate for the \textit{longest} link among all channels). See Appendix J for numerical results.

\section{Conclusion}

In summary, we have proposed a method of effectively compensating for channel asymmetry in MDI-QKD by adjusting the two users' intensities and decoupling the two bases (with the 7-intensity protocol being a highly practical example that works well under finite-size effects). Such a method can drastically increase the scenarios MDI-QKD can be applied to while maintaining good key rate. This study provides a powerful and robust software solution for a scalable and reconfigurable MDI-QKD network.

Our method is also a general result that is not limited to the 7-intensity form, but can in principle be used for e.g. alternative number of decoys, or alternative finite-size analysis models (e.g. joint-bounds analysis, or composable security with Chernoff's bound). It is also potentially applicable to other types of quantum communication protocols, such as Twin-Field QKD \cite{TFQKD}, or MDI quantum digital signature \cite{MDIQDS1,MDIQDS2}, which both use WCP sources and decoy-state analysis. We hope that our proposal can inspire more future work on the study of asymmetric protocols. \\

{\it Notes Added}: After the completion of an earlier version of our manuscript, we have now experimentally implemented our protocol in \cite{asymmmetric_experiment,QCrypt}, thus demonstrating clearly the practicality of our work.

\section*{Acknowledgments}

This work was supported by the Natural Sciences and Engineering Research Council of Canada (NSERC), U.S. Office of Naval Research (ONR), the Fundamental Research Funds for the Central Universities of China, National Natural Science Foundation of China Grants No. 61771443 and China 1000 Young Talents Program. A key inspiration for this project comes from study of free-space MDI-QKD. We thank the collaborators Dr. B Qi and Prof. G Siopsis on helpful discussion and collaborative efforts in implementing free-space MDI-QKD, and thank K. McBryde and S. Hammel from SPAWAR Systems Center, Pacific for kindly providing atmospheric data and for the helpful discussions.

\appendix

\section{Note about Adding Fibre}

\begin{figure}[h]
	\includegraphics[scale=0.35]{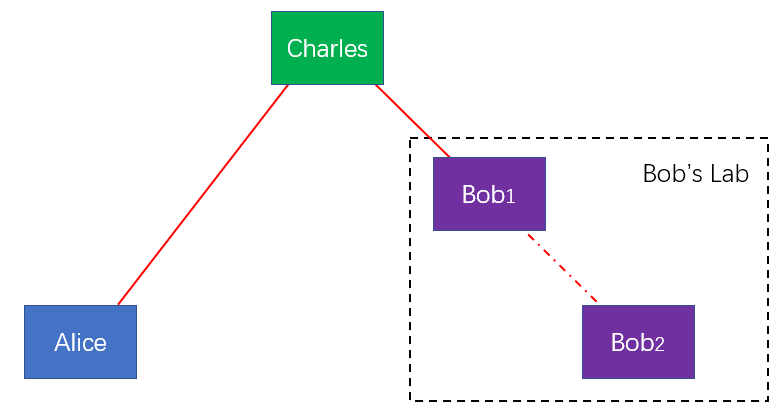}
	\caption{Setup for asymmetric MDI-QKD. When channels are highly asymmetric (e.g. Alice and Bob1), to increase the symmetry in the channel, sometimes one adds additional loss to the system in Bob's lab\cite{mdiPOP}, in exchange for better symmetry. When estimating the key rate, Bob assumes that both Charles-Bob1 and Bob1-Bob2 channels are controlled by Eve. This is therefore a pessimistic estimation of key rate, and is not necessarily the optimal strategy.}
	\label{fig:setup}
\end{figure}

In this section we provide an intuitive description of why adding additional loss is suboptimal, and how our method works better with asymmetric channels.

Previously, when Alice and Bob have asymmetric channels, a common solution is to add fibre (thus adding loss) to the shorter channel in exchange for better symmetry, such as in Ref. \cite{mdiPOP}. Afterwards one selects symmetric intensities for Alice and Bob and acquires higher rate. However, the added fibre lies in Bob's lab, and is in fact securely under control of Bob. But by assuming a symmetric setup, we are effectively relinquishing its control to Eve, and pessimistically estimating the key rate. Therefore, intuitively, this is not necessarily the optimal strategy. We will show with our new protocol that, when the channels are asymmetric, Alice and Bob can independently choose their optimal intensities, and that optimizing intensities and probabilities alone is sufficient to compensate for the different channel losses.

\section{Scaling of Key Rate with Transmittance}

{\color{black}In this section we discuss the scaling properties of key rate versus transmittance, for prior protocols with same parameters for Alice and Bob, and our new protocol that uses different intensities for Alice and Bob. We will show in Appendices B and C that the scaling of the key rate versus distance is mainly determined by the signal states (so long as we have good single photon estimation from decoy states). This also means that, the advantage of our method is really not dependent on the number of decoy states used or the finite-size analysis model used (or lack thereof, in the asymptotic case), and our results are in principle applicable to any protocol that decouples the signal and decoy states in the Z and X bases and allows different intensities for Alice and Bob.}

The transmittance of the two channels are $(\eta_{\text{A}}, \eta_{\text{B}})$, and the asymmetry (mismatch) $x$ is defined as 

\begin{equation}
	x={{\eta_{\text{A}}}\over{\eta_{\text{B}}}}
\end{equation}

\subsection{Single-Photon Source}

Now, let us consider a single-photon case first. That is, suppose Alice and Bob both send perfect single photons only, and the key is generated from two-photon interference. If we ignore the dark counts, the asymptotic key rate can be written as \cite{Preskill}:

\begin{equation}
	R_{SP}=\eta_{\text{A}}\times \eta_{\text{B}} \times [1-2h_2(e_{11})]
\end{equation}

\noindent where $h_2$ is the binary entropy function and $e_{11}$ is the QBER (which is a quantity that, when dark count rate is ignored, is independent of the transmittance). This means that in the perfect single-photon case, the key rate is proportional to $\eta_{\text{A}} \eta_{\text{B}}$, and the mismatch $x$ does not explicitly appear in its expression:

\begin{equation}
	R_{SP} \propto \eta_{\text{A}}\eta_{\text{B}}
\end{equation}

In fact, for a given total distance $L_{\text{A}}+L_{\text{B}}=L$, any positioning of the untrusted relay Charles (e.g. at the midpoint, in Alice's lab, or in Bob's lab) would not affect the key rate, since $\eta_{\text{A}} \eta_{\text{B}}$ only depends on $L$.

\subsection{Weak Coherent Pulse Source}

The previous discussion for single-photon MDI-QKD suggests that, by nature, there is not really any limitation on symmetry for MDI-QKD, at least for the ideal single photon case. Then, where does this dependence of key rate on channel symmetry which we observed come from? In this section, we will show that the scaling of key rate depends on the signal states' trade-off between error-correction and probabilities of sending single-photons, when using WCP sources, rather than privacy amplification (which depends on estimation of single-photon contributions).

More concretely, (as we will prove in the next section) for protocols with symmetric intensities, there are two sharp cut-off values for the mismatch, $x^{max}$ and $x^{min}$, that prevent the protocol from acquiring any key rate when $x>x^{max}$ or $x<x^{min}$ (and optimizing identical intensities $s_{\text{A}}=s_{\text{B}}$ cannot circumvent this problem). This is why protocols such as 4-intensity protocol are limited to near-symmetric positions. 

On the other hand, when a protocol allows independent intensities for Alice and Bob (such as our new 7-intensity protocol described in the main text), we show that the mismatch can always be compensated by optimizing intensities $s_{\text{A}}$ and $s_{\text{B}}$ (hence lifting the limitations $x^{max}$ and $x^{min}$). In fact, we show that for positions with high asymmetry, key rate no longer depends on mismatch $x={{\eta_{\text{A}}}\over{\eta_{\text{B}}}}$ at all, and the optimal key rate only scales with the \textit{smaller} of the two channel transmittances. That is, 

\begin{equation}
	R_{optimal} \propto min(\eta_{\text{A}}^2, \eta_{\text{B}}^2)
\end{equation}

\noindent which means that, the biggest advantage of protocols with independent intensities for Alice and Bob (e.g. 7-intensity protocol) is to completely lift the limitation on channel asymmetry. When compared with adding fibre to maintain asymmetry, we see that its scaling property is still the same, i.e. quadratically related to the (smaller of) channel transmittances, although our method will always perform better (by a constant coefficient) than adding fibre. Moreover, it provides the convenience of not needing additional fibre, which may not be feasible in free-space channels, or when channel mismatch is changing.

Proofs for the above scaling properties can be found in the next section.

\section{Proof of Scaling Properties of Key Rate with Transmittance}

In this section we outline the analytical proofs for the observations on the scaling properties of asymptotic MDI-QKD key rate versus transmittance in the presence of asymmetry, described in Appendix B. We also discuss how the finite-decoy and finite-size effects can be considered as imperfections in the infinite-decoy, infinite-data case, and that the scaling properties are still approximately the same - which are only determined by the signal states' trade-off between error correction and probabilities of sending single photons, and not affected by decoy states.

To simplify the discussion, it is convenient to first use a few crucial approximations as described in Ref.\cite{mdipractical}:\\ 

1. We consider the asymptotic case with infinite data size.

2. We assume an infinite number of decoy states, i.e. Alice and Bob can perfectly estimate the single photon gain $Y_{11}$ and QBER $e_{11}$. In this case, Alice and Bob only need to choose appropriate signal intensities $s_{\text{A}}$, $s_{\text{B}}$.

3. We ignore the dark count rate $Y_0$, when studying the scaling properties with distance (as background noise only affects the maximum transmission distance where transmittance is at the same order as the dark count rate, but does not affect the overall scaling properties of key rate versus distance).

4. When describing the channel model to estimate the observable gain and QBER $Q_{ss}^Z$ and $E_{ss}^Z$ (which affect the error-correction), we make second-order approximations to two functions:

\begin{equation}
	\begin{aligned}
		I_0(x)&\approx 1 + {x^2\over 4} + O(x^4) \\
		e^x &\approx 1+x+{x^2\over 2} + O(x^3)
	\end{aligned}
\end{equation}

\noindent where $I_0$ is the modified bessel function of the first kind. This approximation is relatively accurate when $s_{\text{A}}\eta_{\text{A}}\eta_d$ and $s_{\text{B}}\eta_{\text{B}}\eta_d$ are both small, where $\eta_d$ is the detector efficiency.\\

With the above approximations, one can write the key rate conveniently as (excerpting Eq. C.1 and C.2 from Ref.\cite{mdipractical}):

\begin{equation}
	R = {{\eta_{\text{B}}^2 \eta_d^2}\over 2} G(x,s_{\text{A}},s_{\text{B}})
\end{equation}

\noindent where $G(x,s_{\text{A}},s_{\text{B}})$ is a function determined by $(s_{\text{A}}, s_{\text{B}})$ and the asymmetry $x$ only:

\begin{equation}
	\begin{aligned}
		&G(x,s_{\text{A}},s_{\text{B}}) = x s_{\text{A}} s_{\text{B}} e^{-(s_{\text{A}}+s_{\text{B}})}[1-h_2(e_d-{e_d^2\over 2})] \\
		&- {{2xs_{\text{A}}s_{\text{B}}+(s_{\text{B}}^2+x^2s_{\text{A}}^2)(2e_d-e_d^2)} \over 2} \times f_eh_2(E_{ss}^Z(x,s_{\text{A}},s_{\text{B}})) \\
		&E_{ss}^Z(x,s_{\text{A}},s_{\text{B}}) = {{(s_{\text{B}}+xs_{\text{A}})^2(2e_d-e_d^2)}\over{2[2xs_{\text{A}}s_{\text{B}}+(s_{\text{B}}^2+x^2s_{\text{A}}^2)(2e_d-e_d^2)]}}
	\end{aligned}
\end{equation}

\noindent where $h_2$ is the binary entropy function.

Now, having described the key rate function, we are interested in how it scales with the transmittances $\eta_{\text{A}}$, $\eta_{\text{B}}$, using different optimization strategies for the intensities. We will discuss two cases:\\

1. $R_{symmetric}$, where Alice and Bob use the same intensity $s=s_{\text{A}}=s_{\text{B}}$, and optimize $s$.

2. $R_{optimal}$, where Alice and Bob fully optimize a pair of intensities $s_{\text{A}}, s_{\text{B}}$, which can take different values.\\

\subsection{Symmetrically Optimized Intensities}

Let us consider the case where Alice and Bob use the same intensity $s=s_{\text{A}}=s_{\text{B}}$, and optimize $s$. This is the case discussed by previous protocols (such as the 4-intensity protocol, although here to simplify the proof we focus on infinite-decoy case and only consider signal intensities).

In this case, the function $G$ is optimized over $s$ (and is a function of $x$ only). The rate satisfies

\begin{equation}
	R_{symmetric} = \max\limits_{s}R \propto \eta_{\text{B}}^2 \max\limits_{s}G(x,s,s)
\end{equation}

\noindent therefore, $R_{symmetric}$ is proportional to $\eta_{\text{B}}^2$ when channel mismatch $\eta_{\text{A}} \over \eta_{\text{B}}$ is fixed. 

Moreover, since $R_{symmetric}$ is also proportional to $G(x)$, we will have $R_{symmetric}=0$ if $G(x)=0$. Note that, we can rewrite the signal state QBER $E_{ss}^Z$ as:

\begin{equation}
	\begin{aligned}
		&E_{ss}^Z(x) = {{(1+x)^2(2e_d-e_d^2)}\over{2[2x+(1+x^2)(2e_d-e_d^2)]}}
	\end{aligned}
\end{equation}

\noindent since the equal intensities are canceled out, i.e. $E_{ss}^Z$ is only a function of x. In fact, $E_{ss}^Z$ is a function that minimizes at $x=1$ and reaches $50\%$ (where $R_{symmetric}$ is naturally zero) when $x \rightarrow 0$ or $x\rightarrow \infty$. Therefore, if $G(x)\neq 0$ at $x=1$, there must exist some critical values of $x^{max}$ and $x^{min}$ which result in a sufficiently large QBER such that $G(x)=0$ (and $R_{symmetric}=0$). 

This means that, $R_{symmetric}$ is quadratically related to $\eta_{\text{B}}$ (or $\eta_{\text{A}}$) when mismatch $\eta_{\text{A}} \over \eta_{\text{B}}$ is fixed, but also has two cut-off positions for critical levels of mismatch, beyond which no key can be generated. These two critical mismatch positions are what limit previous MDI-QKD protocols to near-symmetric positions. Also, as we have previously mentioned, we see that this critical dependence on mismatch actually comes from the error-correction part (which involves $E_{ss}^Z$).

\subsection{Fully Optimized Intensities}

\begin{figure}[h]
	\includegraphics[scale=0.4]{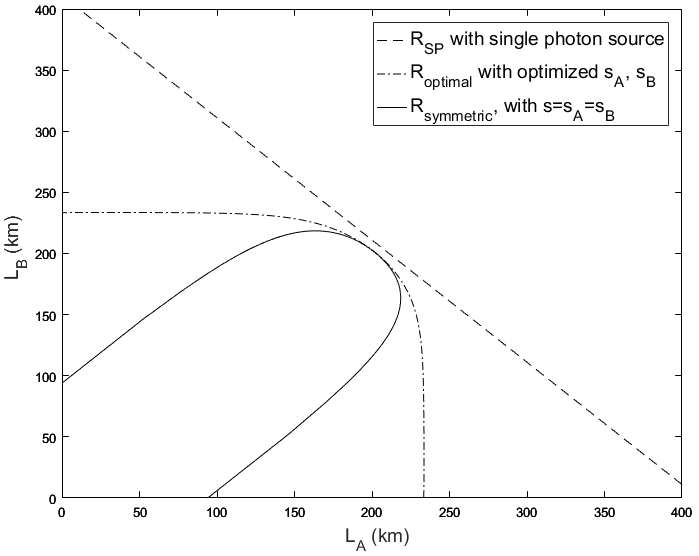}
	\caption{Rate vs distance contours for single photon MDI-QKD $R_{SP}$, decoy-state MDI-QKD with symmetric intensities $R_{symmetric}$, and with fully optimized intensities $R_{optimal}$. We plot the contour line of $R=10^{-9.5}$. Here for a better comparison with WCP sources, we arbitrarily set a probability $P_{11}=s_{\text{A}}s_{\text{B}}\times e^{-(s_{\text{A}}+s_{\text{B}})}$ (where $s_{\text{A}}=s_{\text{B}}=0.6533$) of single photon pairs being sent when calculating $R_{SP}$. For the decoy-state case, as described in Appendix C, we assume infinite decoys, infinite data size, ignore dark count rate, and take second-order approximation when calculating gain and QBER (so that we only focus on the ideal scaling properties of key rate with distance). As can be seen, $R_{SP}$ is not limited by asymmetry, and takes constant value for any fixed $L_{\text{A}}+L_{\text{B}}$ (meaning that the dependence of key rate on asymmetry does not come from single photon contributions in the privacy amplification part when using WCP sources). For decoy-state MDI-QKD, we can clearly see $R_{symmetric}$ being limited by the two cut-off lines where $|L_{\text{A}}-L_{\text{B}}|$ takes maximum value (which corresponds to critical values of channel mismatch $x^{max}$ and $x^{min}$). On the other hand, $R_{optimal}$ is not limited by asymmetry, and has contours nearly perpendicular to the axes when asymmetry is high (meaning that, when one channel is significantly longer than the other, $R_{optimal}$ is only dependent on the longer channel).}
	\label{fig:contours}
\end{figure}

Now, let us consider the case where Alice and Bob are allowed to fully optimize their intensities $s_{\text{A}},s_{\text{B}}$ (such as in the 7-intensity protocol, although again, here we only focus on the signal states in the infinite-decoy case).

In this case, the function $G$ is optimized over $s_{\text{A}}, s_{\text{B}}$. The rate satisfies

\begin{equation}
	R_{optimal} = \max\limits_{s_{\text{A}},s_{\text{B}}}R \propto \eta_{\text{B}}^2 \max\limits_{s_{\text{A}},s_{\text{B}}}G(x,s_{\text{A}},s_{\text{B}})
\end{equation}

Now, let us focus on the properties of $G(x,s_{\text{A}},s_{\text{B}})$. Looking at its expression Eq. (C3) in the previous section, we make the important observation that, except for the term $e^{-(s_{\text{A}}+s_{\text{B}})}$ in the single photon probabilities, every other term is only a function of $s_{\text{B}}$ and $xs_{\text{A}}$ (rather than $x$ and $s_{\text{A}}$ separately). We can re-write $G(x,s_{\text{A}},s_{\text{B}})$ as 

\begin{equation}
	\begin{aligned}
		&G'(x,s_{\text{A}}',s_{\text{B}}) = s_{\text{A}}' s_{\text{B}} e^{-s_{\text{A}}'\over x}e^{-s_{\text{B}}}[1-h_2(e_d-{e_d^2\over 2})] \\
		&- {{2s_{\text{A}}'s_{\text{B}}+(s_{\text{B}}^2+s_{\text{A}}'^2)(2e_d-e_d^2)} \over 2} \times f_eh_2(E_{ss}^Z(s_{\text{A}}',s_{\text{B}})) \\
		&E_{ss}^Z(s_{\text{A}}',s_{\text{B}}) = {{(s_{\text{B}}+s_{\text{A}}')^2(2e_d-e_d^2)}\over{2[2s_{\text{A}}'s_{\text{B}}+(s_{\text{B}}^2+s_{\text{A}}'^2)(2e_d-e_d^2)]}}
	\end{aligned}
\end{equation}

\noindent where we define \textit{equivalent intensity} $s_{\text{A}}'$ as 

\begin{equation}
	s_{\text{A}}'={s_{\text{A}} \times x}
\end{equation}

Moreover, if $\eta_{\text{A}} \gg \eta_{\text{B}}$ (i.e. mismatch $x \gg 1$), we can approximately assume that 

\begin{equation}
	e^{-s_{\text{A}}'\over x} \approx 1
\end{equation}

\noindent which means that we can rewrite $\max\limits_{s_{\text{A}},s_{\text{B}}}G(x,s_{\text{A}},s_{\text{B}})$ as

\begin{equation}
	\begin{aligned}
		G^{max}=\max\limits_{s_{\text{A}}',s_{\text{B}}} &G'(s_{\text{A}}',s_{\text{B}})\\
	\end{aligned}
\end{equation}

\noindent which, importantly, is a constant value \textit{not} dependent on the value of $x$, when $x \gg 1$. The actual value of $s_{\text{A}}$ equals 

\begin{equation}
	s_{\text{A}}={s_{\text{A}}' \over x}
\end{equation}

\noindent Physically, this means that, when there is asymmetry between Alice and Bob's channels, we can compensate for this asymmetry by adjusting the intensities, to keep the same "equivalent intensity" received by Charles and keep $E_{ss}^Z$ at a low value. In this case, $E_{ss}^Z$ is no longer limited by the mismatch $x$, and we can perform MDI-QKD at arbitrary values of asymmetry. 

Also, the key rate is now given by:

\begin{equation}
	R_{optimal} \propto \eta_{\text{B}}^2 G^{max}
\end{equation}

\noindent This means that, when $\eta_{\text{A}} \gg \eta_{\text{B}}$ (e.g. the "single-arm" case previously mentioned where $L_{\text{A}}$ is much shorter than $L_{\text{B}}$), the key rate of asymmetric MDI-QKD is only related to $\eta_{\text{B}}$ and still quadratically scales with $\eta_{\text{B}}$. When $\eta_{\text{B}} \gg \eta_{\text{A}}$, though, we can rewrite $x'={\eta_{\text{B}}\over \eta_{\text{A}}}$, and rewrite

\begin{equation}
	R_{optimal} \propto \eta_{\text{A}}^2 \max\limits_{s_{\text{B}}',s_{\text{A}}} G'(s_{\text{B}}',s_{\text{A}})
\end{equation}

\noindent Therefore, overall, 

\begin{equation}
	R_{optimal} \propto min(\eta_{\text{A}}^2, \eta_{\text{B}}^2)
\end{equation}

Now, we plot the two cases (symmetric intensities and fully optimized intensities) in a contour plot. As we can observe in Fig.\ref{fig:contours}, the key rate $R_{symmetric}$ has two cut-off mismatch positions beyond which key rate is zero. This limitation is removed when full optimization of intensities is implemented. Moreover, for $R_{optimal}$, we see that the contours are perpendicular to the axes in high asymmetry regions, which means that the key rate only scales with the longer of the two channels.\\

{\color{black}
Also, note that, from Eqs. (C4), (C6), we can also make the observation that there is never any need to add fibre to the shorter channel when fully optimizing the intensities, and our new method always provides higher key rate than prior art technique of adding fibre till channels are symmetric, while using same intensities for Alice and Bob. 

To show this, consider the system having a fixed longer channel $L_{\text{B}}$ (i.e. suppose $\eta_{\text{B}}$ is fixed and $\eta_{\text{A}} > \eta_{\text{B}}$, $x ={\eta_{\text{A}} \over \eta_{\text{B}}} > 1$). Adding loss to $\eta_{\text{A}}$ is equivalent to decreasing $x$.

With symmetric intensities (and adding loss till $\eta_{\text{A}}=\eta_{\text{B}}$), the key rate can be written as:

\begin{equation}
R_{symmetric} = {{\eta_d^2  \eta_{\text{B}}^2 } \over 2} \max\limits_{s}G(1,s,s)
\end{equation}

Suppose we fully optimize the intensities for this case with added fibre, we will obtain the same key rate (since for $x=1$, i.e. symmetric setup, the optimal choice of intensities satisfies $s_{\text{A}}=s_{\text{B}}$):

\begin{equation}
\max\limits_{s}G(1,s,s) = \max\limits_{s_{\text{A}},s_{\text{B}}}G(1,s_{\text{A}},s_{\text{B}})
\end{equation}

However, let us compare it with the case of using fully optimized intensities and no additional loss:

\begin{equation}
R_{optimal} = {{\eta_d^2  \eta_{\text{B}}^2 } \over 2} \max\limits_{s_{\text{A}},s_{\text{B}}}G(x,s_{\text{A}},s_{\text{B}})
\end{equation}

As described in Eq. (C7), we can re-write $G(x,s_{\text{A}},s_{\text{B}})$ as $G'(x,s_{\text{A}}',s_{\text{B}})$ (recall that the equivalent intensity $s_{\text{A}}'$ is defined as $xs_{\text{A}}$). We make the observation that $G'(x,s_{\text{A}}',s_{\text{B}})$ strictly increases with $x$. That is, for any two given values of $s_{\text{A}}',s_{\text{B}}$ and $x>1$,

\begin{equation}
G'(x,s_{\text{A}}',s_{\text{B}}) > G'(1,s_{\text{A}}',s_{\text{B}})
\end{equation}

\noindent hence after optimization we also have 

\begin{equation}
\max\limits_{s_{\text{A}}',s_{\text{B}}}G'(x,s_{\text{A}}',s_{\text{B}}) > \max\limits_{s_{\text{A}}',s_{\text{B}}}G'(1,s_{\text{A}}',s_{\text{B}})
\end{equation}

\noindent which means that, when fully optimizing Alice and Bob's intensities (which already compensate for the mismatch between channels), it is always optimal not to add any additional loss to the channels. Moreover, combining Eqs. (C15), (C16), (C17), (C19), we can see that

\begin{equation}
\begin{aligned}
R_{optimal} &= {{\eta_d^2  \eta_{\text{B}}^2 } \over 2} \max\limits_{s_{\text{A}},s_{\text{B}}}G(x,s_{\text{A}},s_{\text{B}})\\ 
&> {{\eta_d^2  \eta_{\text{B}}^2 } \over 2} \max\limits_{s}G(1,s,s) = R_{symmetric}
\end{aligned}
\end{equation}

That is, compared to the case where one adds loss to $\eta_{\text{A}}$ until $\eta_{\text{A}}=\eta_{\text{B}}$, our new protocol always provides higher key rate as long as the channels are asymmetric. Intuitively, this is because adding fibre while using same intensities for Alice and Bob is in fact an suboptimal subset of the overall set of strategies Alice and Bob can take (which includes adjusting Alice and Bob’s intensities independently, as well as adding any length of fibres to change $x$). Even when considering adding fibre as one of the valid variables, we have shown that optimal point always happens when no fibre is added. Therefore, our method is a better optimized strategy than adding fibre because it considers a larger parameter space.

Note that, fully optimizing Alice and Bob's intensities does not change the fundamental scaling property - the key rate is still \textit{quadratically} related to transmittance in the longer arm - However, it always provides better key rate than prior art techniques, and also offers the great convenience of not having to physically add loss to the channels and being able to implement everything in software.
}
\subsection{Practical Imperfections}

\begin{figure}[h]
	\includegraphics[scale=0.38]{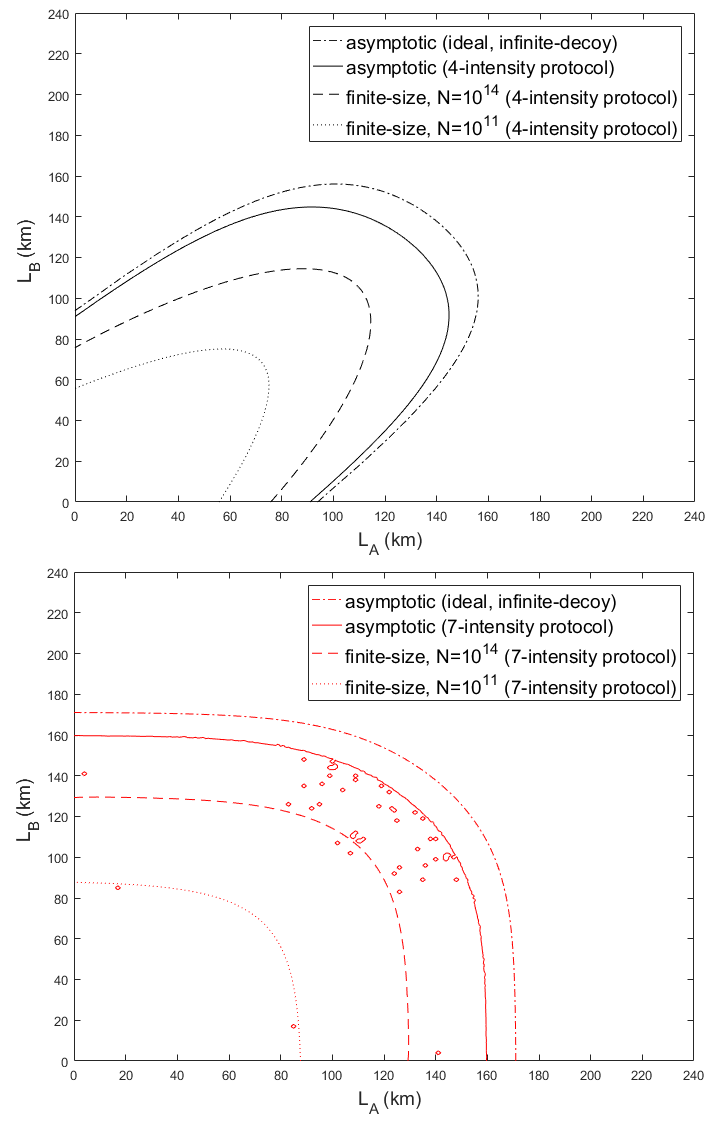}
	\caption{Contours of rate vs distance for decoy-state MDI-QKD, under different assumptions for practical imperfections, for the key rates for asymptotic case with infinite decoys (and ideal assumption of zero dark count rate and 100\% detector efficiency), asymptotic case with 4-intensity/7-intensity protocol (with practical device parameters), and finite-size case with 4-intensity/7-intensity protocol. Top: protocols with identical intensities for Alice and Bob, Bottom: protocols with fully optimized intensities. (Note that in the bottom plot there are some noises in the asymptotic 7-intensity protocol key rate. This is because the optimal $\nu$ can take a very small value in the ideal case where data size is infinitely large. This results in some numerical noises in computer simulations). We plot the contour lines of $R=10^{-7}$. As can be observed here, the finite number of decoys, the non-ideal experimental parameters, and the finite-size effects are all imperfections that reduce the key rate. However, the overall shapes of the contours still remain largely the same, which follow the upper bounds given by the ideal infinite-decoy case. (Except for 4-intensity protocol under finite-size effect, which no longer has two clear cut-off mismatch positions, but is still severely limited by channel asymmetry, while 7-intensity protocol lifts this constraint completely).}
	\label{fig:contours_compare}
\end{figure}

Up to here we have analytically shown how choosing to fully optimize the intensities can affect the key rate, for the asymptotic, infinite-decoy case. The behavior of contours as shown in Fig.\ref{fig:contours} is a result of $s_{\text{A}},s_{\text{B}}$ compensating for the difference in channel loss. However, we have so far assumed perfect knowledge of single-photon contributions, and have not yet discussed the decoy-state intensities. Moreover, non-ideal experimental parameters (including dark count rate and detector efficiency), and finite-size effects will both affect the key rate. Here in this subsection, we compare the key rate under more practical assumptions, and show that the above factors can be considered as \textit{imperfections} that reduce the key rate, but maintain similar contour shapes and scaling properties for the key rate - that is, we will still observe a high dependence on asymmetry for protocols with identical intensities for Alice and Bob, and fully optimizing intensities can completely lift this limitation.

In practice, with a finite number of decoys (for instance, for 4-intensity and 7-intensity protocols, where Alice and Bob choose respectively three decoy intensities, $\mu,\nu,\omega$), the estimation of $Y_{11}$ and $e_{11}$ is not perfect, therefore the key rate will be slightly lower than the aforementioned infinite-decoy case. Moreover, to accurately estimate $Y_{11}$ and $e_{11}$, the decoy intensities need to be optimized to compensate for channel loss, too. As described in Subsection II.A in the main text, the decoy states should maintain balanced arriving intensities at Charles (e.g. $\mu_{\text{A}}\eta_{\text{A}}=\mu_{\text{B}}\eta_{\text{B}}$), to ensure good HOM visibility and low QBER in the X basis. Note that, the optimization of decoy intensities has a very different purpose from that of the signal intensities $s_{\text{A}},s_{\text{B}}$ - the signal intensities are optimized so as to reduce $E_{ss}^Z$ (while keeping single photon probability $s_{\text{A}}s_{\text{B}}e^{-(s_{\text{A}}+s_{\text{B}})}$ high) and maximize the key rate, while the decoy intensities are optimized to estimate $Y_{11}^{L}$ and $e_{11}^{U}$ as accurately as possible, whose ideal values $Y_{11}$ and $e_{11}$ (used in the infinite-decoy case above) provide an upper bound for the practical key rate with finite number of decoys. As we see in Fig.\ref{fig:contours_compare}, the asymptotic key rate with a finite number of decoys follows a similar shape as its upper bound, the infinite-decoy case. 

Additionally, the detector efficiency (which is equivalent to channel loss) contributes to a uniformly shifted key rate in both $L_{\text{A}}$ and $L_{\text{B}}$ directions, while dark counts reduce the key rate more significantly in the higher loss region (both of which we have ignored in the ideal case as described at the beginning of this section). However, as observed in Fig.\ref{fig:contours_compare} (the solid lines consider both finite-decoys and practical parameters), these factors do not change the overall shape of the contours either.

Lastly, finite-size effect will reduce the key rate significantly. As observed in Fig.\ref{fig:contours_compare} bottom plot, while the key rate is reduced, the contour shapes remain largely unchanged (meaning that even under finite-size effect, the 7-intensity protocol can still effectively compensate for channel asymmetry effectively). In Fig.\ref{fig:contours_compare} top plot, we can find similar observations, that finite-size effect reduces the overall key rate. However, note that, under finite-size effect, the shapes of key rate contours for the 4-intensity protocol are somewhat different, and no longer follow the two cut-off positions $x^{upper}$, $x^{lower}$ for channel mismatch (which appear as straight lines in e.g. Fig.\ref{fig:contours}). This is because, though the key rate is still limited by $E_{ss}^Z$ (which causes the cut-off mismatch positions), it is also limited by the estimation of $Y_{11}^{L}$ and $e_{11}^{U}$ using the decoy states. Compared to the asymptotic case, here under finite-size effect, the increased $e_{11}^U$ is likely a more severe limiting factor than $E_{ss}^Z$, and not being able to choose independent intensities for Alice and Bob prevents an accurate estimation of $Y_{11}^{L}$ and $e_{11}^{U}$ (due to poor HOM visibility in X basis caused by unbalanced intensities). Therefore, here the dependence of key rate on channel asymmetry is present in both privacy amplification and error-correction terms, and the shapes of contours are a result of both effects. (The difference in contour shape from the infinite-decoy case is more prominent for finite-size case, likely because the key rate is more sensitive to $e_{11}^{U}$ here). Importantly, under finite-size effects, the key rate for 4-intensity protocol is still highly limited by channel asymmetry, while 7-intensity protocol completely removes such a constraint and allows two channels with arbitrary asymmetry between them.

\section{Note about Decoupling Signal and Decoy Intensities}

In this section we provide a simple intuitive explanation for why our protocol provides a better choice of decoy and signal intensities. 

\subsection{Performance}

Let us recall again the key rate formula of MDI-QKD \cite{mdiqkd,mdifourintensity}:
\begin{equation}
	\begin{aligned}
		R=P_{s_{\text{A}}}P_{s_{\text{B}}} \{(s_{\text{A}} e^{-s_{\text{A}}})(s_{\text{B}} e^{-s_{\text{B}}}) Y_{11}^{X,L}[1-h_2(e_{11}^{X,U})]\\
		-f_eQ_{ss}^Z h_2(E_{ss}^Z)\}
	\end{aligned}
\end{equation}

Here there are three criteria that determine whether a MDI-QKD protocol generates good key rate in the presence of channel asymmetry:\\

(a) Similar arriving intensities at Charles in the X basis, in order to have good HOM interference and keep QBER low in the X basis (which is important for a good estimation of $e_{11}^{X,U}$).

(b) Similar arriving intensities at Charles in the Z basis, in order to keep QBER $E_{ss}^Z$ low in the Z basis (which is due to misalignment), although this term is much less sensitive to difference in intensities than (a).

(c) A high enough probability of sending single-photons, $s_{\text{A}} e^{-s_{\text{A}}}s_{\text{B}} e^{-s_{\text{B}}}$. Note that both criteria (b) and (c) involve the signal states $s_{\text{A}}$, $s_{\text{B}}$, so there is a trade-off between (b) and (c).\\

Prior protocols require Alice and Bob to use the same set of intensities. This overly constrains the solution space (because Alice and Bob try to use the same set of intensities to satisfy (a), (b) and (c) simultaneously), and leaves high QBER in both the X and Z bases, and thus resulting in low key rate when channels are asymmetric. 

By relaxing this constraint (allowing Alice and Bob to have different intensities), and decoupling criteria (a) from criteria (b) and (c) by allowing independent decoy and signal intensities, we can satisfy (a) nicely, while simultaneously achieving a good trade-off between (b) and (c), hence ensuring a high key rate.

\textit{\textbf{Remark}}: for more detail on the trade-off between (b) and (c), here (b) is optimal when arriving intensities are matched, i.e. $s_{\text{A}}/s_{\text{B}}=\eta_{\text{B}}/\eta_{\text{A}}$, and (c) is independent of asymmetry and is optimal when signal intensities are both 1. In fact, since $E^{Z}_{ss}$ is much less sensitive to $s_{\text{A}}/ s_{\text{B}}$, such a trade-off between two terms favors (c) more than (b), thus the optimal $s_{\text{A}}/s_{\text{B}}$ is often closer to 1 than $\eta_{\text{B}}/\eta_{\text{A}}$. The actual optimal signal intensities can be found by numerical optimization, as described in Section II.B. An example of $\mu_{\text{A}}/\mu_{\text{B}}$ and $s_{\text{A}}/s_{\text{B}}$ can also be seen in Fig.\ref{fig:ratio}, where we observe that  $\mu_{\text{A}}/\mu_{\text{B}}$ follows $\eta_{\text{B}}/\eta_{\text{A}}$ rather closely, while $s_{\text{A}}/ s_{\text{B}}$ has relatively much more freedom in its optimization (between $1$ and $\eta_{\text{B}}/\eta_{\text{A}}$).\\

\subsection{Security}
{\color{black}
Another point is, our parameter choice not only provides better performance, but also ensures no less security than in prior art protocols. The security of our protocol relies on two key assumptions: (1) Given the same photon number n in a pulse, Eve has no way of differentiating the decoy states from signal states in the same basis, and (2) the single photons pairs in X and Z bases cannot be distinguished from each other.

Under the first assumption, the yields of photon numbers $m, n$ in Alice’s and Bob’s channel in the X basis satisfy $Y^X_{m,n} (\mu_A^1,\mu_B^1 )=Y^X_{m,n} (\mu_A^2,\mu_B^2)$, when Alice and Bob use two different intensity pairs $\{\mu_A^1,\mu_B^1\}$ and $\{\mu_A^2,\mu_B^2\}$ ($\mu_A^1,\mu_A^2$ can be any state among Alice's decoy intensities, and similarly $\mu_B^1,\mu_B^2$ can be any state among Bob's decoy intensities). Similarly, the QBERs satisfy $e^X_{m,n} (\mu_A^1,\mu_B^1 )=e^X_{m,n} (\mu_A^2,\mu_B^2)$. This is a reasonable assumption because, by only observing the photon numbers in a pulse, Eve has no way of telling which intensity setting it comes from. Therefore, the observables from different intensity combinations can be used as linear constraints for decoy-state analysis in the X basis. This idea is in essence the same as the foundations for decoy-state BB84 in Refs. \cite{decoystate_Hwang,decoystate_LMC,decoystate_Wang} and decoy-state MDI-QKD in Ref. \cite{mdiqkd}. Note that, for successful decoy-state analysis we do not require the symmetry between the two bases, i.e. $Y^X_{m,n} (\mu_A^1,\mu_B^1 )=Y^Z_{m,n} (\mu_A^2,\mu_B^2)$ or $e^X_{m,n} (\mu_A^1,\mu_B^1 )=e^Z_{m,n} (\mu_A^2,\mu_B^2)$ for multi-photon pulses are not required.

The second assumption is that single photon pairs sent in X or Z bases cannot be distinguished from each other. That is because, regardless of the basis used for encoding, the single photon pairs that can trigger Charles’ detection events are always in Bell states $\ket{\psi^+}$ and $\ket{\psi^-}$. That is, the states sent by Alice and Bob satisfy $\rho^X_{1,1}=\rho^Z_{1,1}$, and Eve has no way of telling apart which basis a pair of single photons come from. Therefore, we can safely assume that $Y_{11}^X=Y_{11}^Z$, which is the reason we can perform decoy-state in the X basis only to estimate $Y_{11}^X$, and assume that $Y_{11}^Z=Y_{11}^X$. 

The security of a scheme of decoupling the bases in MDI-QKD and using the assumption of $Y_{11}^Z=Y_{11}^X$ has been theoretically studied in Ref. \cite{mdifourintensity} and (in the Appendix of) Ref. \cite{mdiToshiba} \footnote{{\color{black}The idea of decoupling the bases was first studied for BB84 in Ref. \cite{biasedQKD}. There the assumption was $Y_{1}^Z=Y_{1}^X$ for single photons instead of single photon pairs.}}, and the scheme has also been experimentally demonstrated in Ref. \cite{mdi404km} and Ref. \cite{mdiToshiba} - although all these works were focused on the scenario of symmetric channels only, and did not discuss the role of decoupled bases in compensating channel asymmetry, which is one of the main novelties of our work. However, physically, the only difference between the signals sent from Alice and Bob in our protocol and those in prior protocols will be the different intensities on the two arms (which we know, from assumption one $Y^X_{m,n} (\mu_A^1,\mu_B^1 )=Y^X_{m,n} (\mu_A^2,\mu_B^2)$, will not affect the security of decoy-state analysis), and for decoupled bases we make the same assumption $Y_{11}^Z=Y_{11}^X$ about single photons, which is no less secure than prior works either.
}

\section{Generality of Our Method: MDI-QKD Protocols other than Three Decoy States}

{\color{black}
	
	In the main text we have focused on the 7-intensity protocol, where Alice and Bob each uses one signal intensity $s_A$ ($s_B$), and three decoy intensities $\mu_A,\nu_A, \omega$ ($\mu_B,\nu_B, \omega$). However, the core idea of our protocol lies in two key points: (1) X and Z bases are decoupled, where decoy-states in the X basis bound Eve's information and signal state in the Z basis encodes the key, and (2) Alice and Bob use different intensities to compensate for channel asymmetry. This means that our protocol is not limited to the 7-intensity protocol, but can easily be applied to other protocols too, as long as points (1) and (2) are satisfied. 
	
	In this section, we demonstrate the generality of our method by actually applying it to other kinds of MDI-QKD protocols where Alice and Bob use a different number of decoy intensities in the X basis, and show that similar advantages as with the 7-intensity protocol can be observed when using asymmetric intensities. We will also show with numerical results that, although these alternative protocols will certainly work, the 7-intensity protocol provides a good balance between performance and ease of experimental implementation.\\
	
	Here we compare three cases:
	
	1. Alice and Bob each uses two decoy intensities $\mu_A,\nu_A$ ($\mu_B,\nu_B$) in the X basis. We denote this case as a \textbf{6-intensity protocol} (including the two signal intensities), where the parameter choices are: 
	
	\begin{equation}
		\begin{aligned}
			&[s_A,\mu_A,\nu_A,P_{s_A},P_{\mu_A},P_{\nu_A},\\
			&s_B,\mu_B,\nu_B,P_{s_B},P_{\mu_B},P_{\nu_B}]
		\end{aligned}
	\end{equation}

	\noindent Here $P_{\nu_A}=1-P_{s_A}-P_{\mu_A}$ and $P_{\nu_B}=1-P_{s_B}-P_{\mu_B}$. This is similar to the "one-decoy" setup that was discussed in Ref. \cite{mdiparameter}. Note that here it's not a ``5-intensity" protocol, because using $\mu_A,\mu_B,\omega$ alone is not sufficient to satisfactorily bound the single-photon contributions and will result in low or zero key rate. Therefore, in this setup, the vacuum state is not used, and Alice and Bob each uses two non-zero decoy states.\\
	
	2. Alice and Bob each uses three decoy intensities $\mu_A,\nu_A, \omega$ ($\mu_B,\nu_B, \omega$) in the X basis. This case is the \textbf{7-intensity protocol} we discussed in the main text, where the parameter choices are 
	
	\begin{equation}
		\begin{aligned}
			&[s_A,\mu_A,\nu_A,\omega,P_{s_A},P_{\mu_A},P_{\nu_A},P_{\omega_A},\\
			&s_B,\mu_B,\nu_B,\omega,P_{s_B},P_{\mu_B},P_{\nu_B},P_{\omega_B}]
		\end{aligned}
	\end{equation}
	
	\noindent Here $P_{\omega_A}=1-P_{s_A}-P_{\mu_A}-P_{\nu_A},P_{\omega_B}=1-P_{s_B}-P_{\mu_B}-P_{\nu_B}$, and $\omega$ is the vacuum state (for simplicity we can assume $\omega=0$).\\
	
	3. Alice and Bob each uses four decoy intensities $\mu_A,\nu_A, \nu_{2A}, \omega$ ($\mu_B,\nu_B, \nu_{2B}, \omega$) in the X basis. We denote this case as a \textbf{9-intensity protocol}, where the parameter choices are 
	
	\begin{equation}
		\begin{aligned}
			&[s_A,\mu_A,\nu_A,\nu_{2A}, \omega, P_{s_A},P_{\mu_A},P_{\nu_A}, P_{\nu_{2A}},P_{\omega_A},\\
			&s_B,\mu_B,\nu_B,\nu_{2B}, \omega, P_{s_B},P_{\mu_B},P_{\nu_B}, P_{\nu_{2B}},P_{\omega_B}]
		\end{aligned}
	\end{equation}
	
	\noindent Here $P_{\omega_A}=1-P_{s_A}-P_{\mu_A}-P_{\nu_A}-P_{\nu_{2A}},P_{\omega_B}=1-P_{s_B}-P_{\mu_B}-P_{\nu_B}-P_{\nu_{2B}}$, and $\omega$ is the vacuum state.
	
	Note that in all of these three protocols, the key rate formula stays the same as Eq. (1):
	
	\begin{equation}
		\begin{aligned}
			R=P_{s_{\text{A}}}P_{s_{\text{B}}} \{(s_{\text{A}} e^{-s_{\text{A}}})(s_{\text{B}} e^{-s_{\text{B}}}) Y_{11}^{X,L}[1-h_2(e_{11}^{X,U})]\\
			-f_eQ_{ss}^Z h_2(E_{ss}^Z)\}
		\end{aligned}
	\end{equation}
	
	\noindent What is changing here is only the estimation of the single-photon contributions, namely the yield $Y_{11}^{X,L}$ and QBER $e_{11}^{X,U}$. While we have analytical bounds for decoy-state analysis \cite{mdipractical} for the 7-intensity protocol, we use a linear programming approach to numerically estimate $Y_{11}^{X,L}$ and QBER $e_{11}^{X,U}$ in the 6-intensity and 9-intensity cases. Such an approach has been widely discussed in literature as in Refs.\cite{MDIAnalytical,mdiparameter,mdiChernoff}.

	Now, we perform numerical simulations with the 6-intensity, 7-intensity and 9-intensity protocols, and show that they all have much higher performance than their symmetric-intensity counterparts when channel asymmetry is present. This demonstrates the generality of our method as using asymmetric intensities can always improve the performance of MDI-QKD with asymmetric channels.
	
	We also compare the performances of the three protocols with each other, and show that using more decoy intensities can always guarantee higher or equal performance than using fewer decoy intensities, regardless of data size and asymmetry. The 7-intensity always provides no smaller key rate than the 6-intensity protocol, and although 9-intensity protocol can potentially provide even higher key rate, the advantage is small, and the 7-intensity protocol we used in the main text is a good balance between key rate performance and ease of experimental implementation.
	
	Interestingly, as observed in Fig. \ref{fig:generality_rate} (a)(b), for the 6-intensity and 9-intensity protocols, although the yield $Y_{11}^{X,L}$ and QBER $e_{11}^{X,U}$ are estimated numerically using linear programming, there is still a "ridge" (discontinuity in first-order derivatives) along ${\mu_A \over \mu_B} = {\nu_A \over \nu_B}$, and  ${\nu_A \over \nu_B} = {\nu_{2_A} \over \nu_{2_B}}$ as we saw for 7-intensity protocol in the main text. For 6-intensity protocol, the ridge is very clearly shown. For 9-intensity protocol, the ridge exists but is less prominent, and sometimes not visible (likely because, e.g. if two pairs of proportional decoy states ${\nu_A \over \nu_B} = {\nu_{2_A} \over \nu_{2_B}}$ already provide good estimation of $Y_{11}^{X,L},e_{11}^{X,U}$, the third pair $\mu_A,\mu_B$ has more freedom, and wouldn't affect the decoy-state analysis or the key rate too much even if it doesn't provide good HOM visibility and results in high $E_{\mu\mu}^X$).

	\begin{figure}[h]
		{\color{black}
			\includegraphics[scale=0.128]{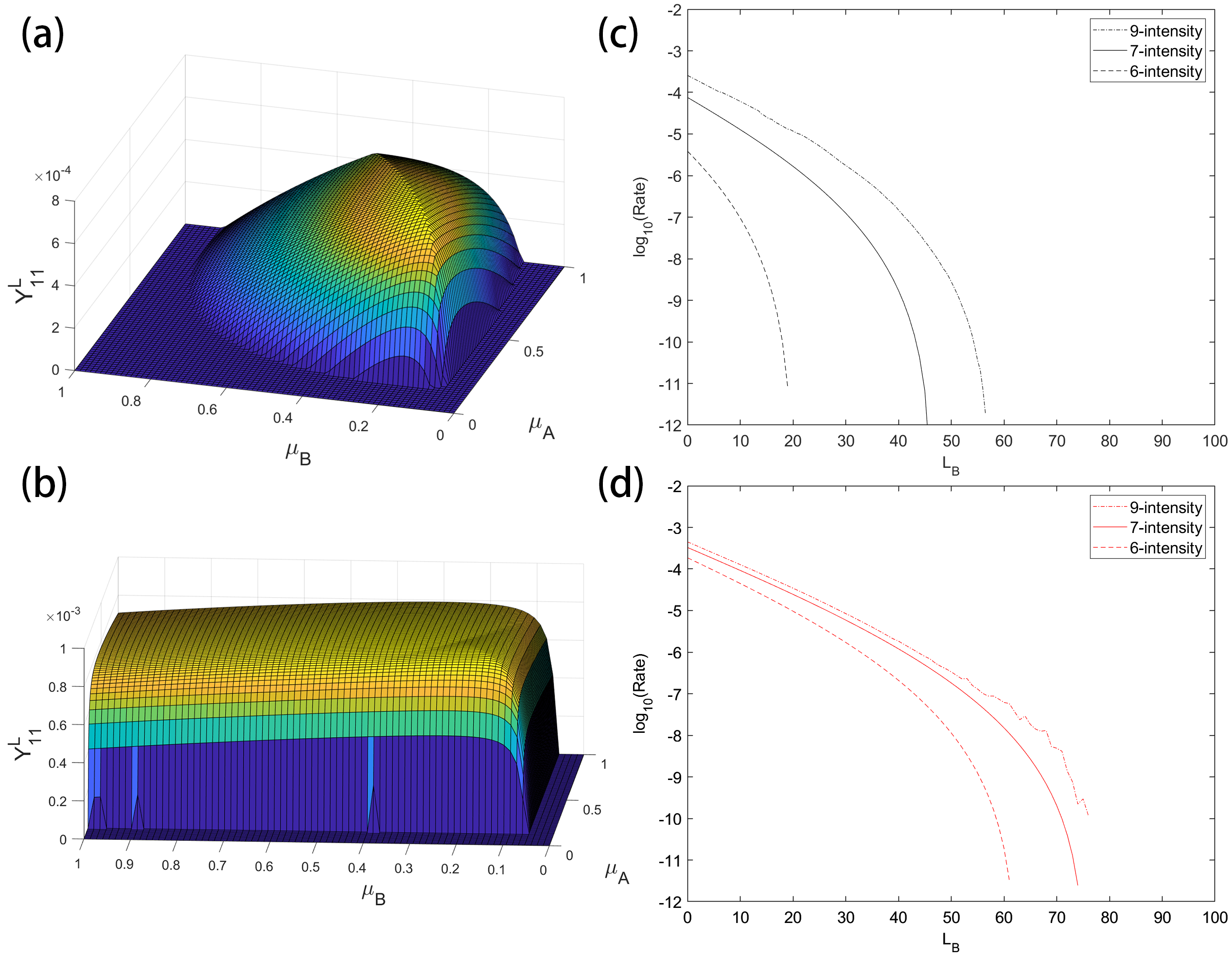}
			\caption{Left: Examples of $Y_{11}^{X,L}$ versus $\mu_A$ and $\mu_B$ where other parameters are all fixed for (a) 6-intensity protocol and (b) 9-intensity protocol. Just like for 7-intensity protocol, we can see a ridge along ${\mu_A \over \mu_B} = {\nu_A \over \nu_B}$, or ${\mu_A \over \mu_B} = {\nu_A \over \nu_B} = {\nu_{2_A} \over \nu_{2_B}}$. Note that the ridge is a lot less obvious for 9-intensity protocol (and sometimes is not visible) likely because two proportional pairs of decoy-states can estimate single-photon contribution reasonably well, so the third pair ($\mu_A,\mu_B$) here has more freedom in choice of intensities. Right: Comparison of rate vs $L_{\text{B}}$ for 6-intensity, 7-intensity and 9-intensity protocols, where mismatch $x=0.1$, i.e. $L_{A} = L_{B} + 50km$ (assuming fibre loss $0.2dB/km$). The rates are plotted in log-scale. We use the parameters from Table I, and $N=10^{12}$. (c) using symmetric intensities for Alice and Bob, (d) using fully optimized asymmetric parameters for Alice and Bob. As can be seen, using asymmetric intensities can greatly improve key rate for all three protocols, when channel asymmetry is present. Note that there is a higher amount of noise present for the 9-intensity case due to the numerical instability brought by linear program solvers (similar to that of joint-bound finite-size analysis, which will be discussed in Appendix H), but the key points here are that the 9-intensity protocol also benefits considerably from using asymmetric intensities, and that the 9-intensity protocol does not have a significant advantage over the 7-intensity protocol despite being more complex to implement.}
			\label{fig:generality_rate}
		}
	\end{figure}

	We plot the simulated key rate for the protocols in Fig. \ref{fig:generality_rate} (c)(d). We first consider a similar scenario as Fig. \ref{fig:2d_Results} (c)(d), using parameters from Table I, a channel mismatch of ${\eta_A \over \eta_B} = x = 0.1$, and data size of $N=10^{12}$ (here we use a larger data size than in the main text, since for $N=10^{11}$, 6-intensity protocol with symmetric intensities cannot generate key rate even at $L_B=0km$ so a comparison is not immediately clear in the plot). As shown in Fig. \ref{fig:generality_rate}, for each protocol, allowing asymmetric intensities provides a much higher key rate than using symmetric intensities only, demonstrating the general effectiveness of our method for different protocols under channel asymmetry.
	
	We also make an important observation here: The more decoy intensities one uses, the higher key rate one can obtain after parameter optimization, even with finite-size effects considered (e.g. the 9-intensity protocol always has higher key rate than 7-intensity, and 7-intensity also always has higher rate than 6-intensity). This is because, for instance, the 6-intensity protocol can in fact be considered as a special case of the 7-intensity protocol, just with $P_{\omega_A}$ and $P_{\omega_B}$ infinitely close to zero, and with 9 instead of 4 constraints for e.g. the gains $Q_{ij}^X$ when estimating $Y_{11}^{X,L}$. With close to zero data, the 5 new constraints are obviously very loose (with very large finite size fluctuation) and will not provide any useful information, but the key point is, in a linear program these loose constraints \textit{will not decrease} the key rate. Therefore, any optimal set of parameters for the 6-intensity protocol can also be considered as a valid set of parameters for the 7-intensity protocol, i.e. the parameter space of 6-intensity protocol is a subset of that of the 7-intensity protocol, and the latter always provides \textit{no smaller} key rate than the former (and often the 7-intensity protocol can find a better parameter set in the larger parameter space, resulting in higher key rate).
	
	The same goes for the 9-intensity protocol, but as we have seen in Fig. \ref{fig:generality_rate}, the advantage it provides over the 7-intensity protocol is rather small (compared to e.g. 6-intensity versus 7-intensity), while requiring more complex control of the intensity modulators in the experimental setup, and more complicated data collection and analysis: the users need to collect 16 sets of gains and error-gains, and the parameter optimization is also a lot slower and more unstable (evaluating the linear program is on average slower than analytical expression by about 50 times, and linear programs also introduce numerical instabilities). Similar observations have been made for the symmetric case in Ref. \cite{mdiparameter} (although in this paper the signal states are not decoupled from decoy states so the protocols are slightly different) that using decoy states $\{\mu,\nu,\omega\}$ provides higher key rate than $\{\mu,\nu\}$, but adding one more decoy-state $\nu_2$ provides little further advantage. 
	
	Therefore, our conclusion is that, while our method of asymmetric intensities and decoupled bases surely works well with other protocols such as 6-intensity and 9-intensity protocols, the 7-intensity protocol we introduced in the main text strikes a good balance between key rate performance and the ease of both experimental implementation and data analysis.
}

\section{Decoy State Intensities}

In this section we will described Theorems I and II in more detail, and show their theoretical proofs in the asymptotic limit of infinite data size (Moreover, numerically, we found that Theorems I and II in fact hold true even under finite-size effects).

\subsection{Symmetry of Optimal Decoy Intensities}

\begin{figure*}[t]
	\includegraphics[scale=0.31]{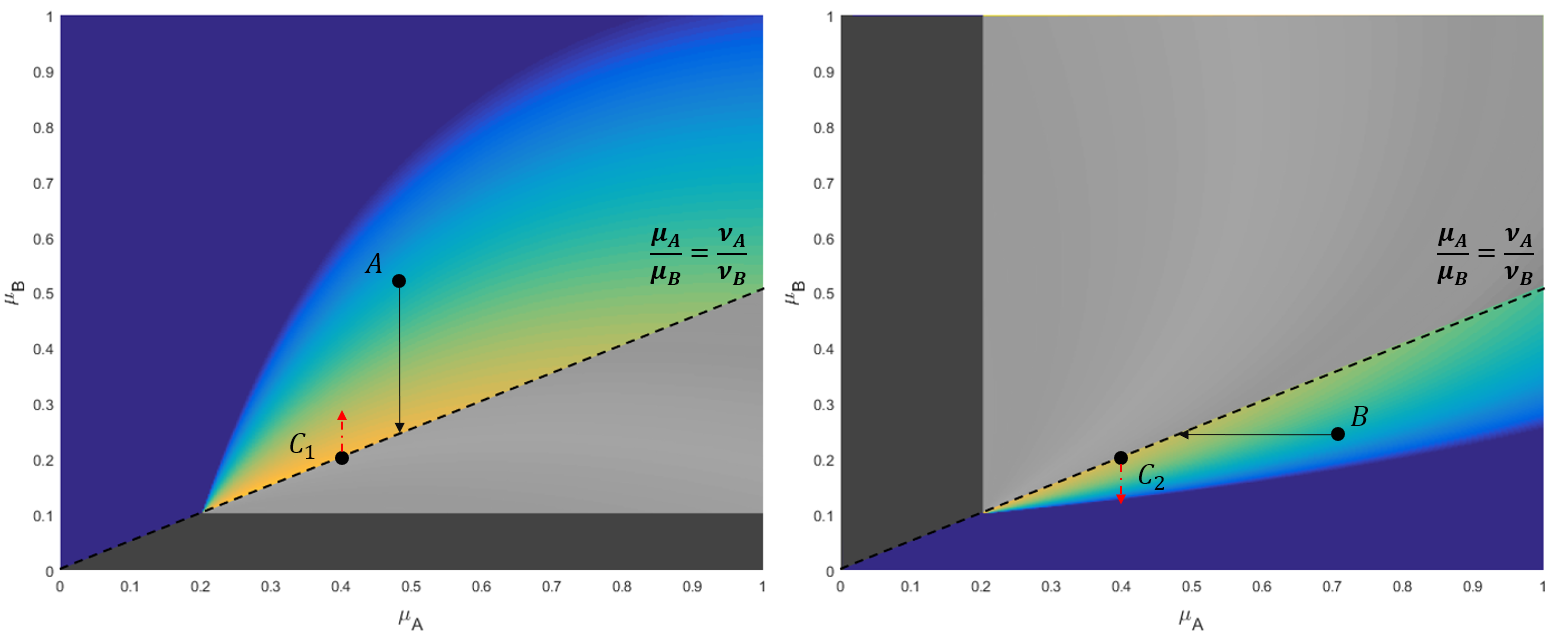}
	\caption{An example of the two difference cases of $Y_{11}^{X,L}$ vs $\mu_{\text{A}}, \mu_{\text{B}}$ function, for fixed values of $\nu_{\text{A}}=0.2$, $\nu_{\text{B}}=0.1$. Left: $Y_{11}^a$, where ${\mu_{\text{A}} \over \mu_{\text{B}}} \leq {\nu_{\text{A}} \over \nu_{\text{B}}}$ ("case 1"); Right: $Y_{11}^b$, where ${\mu_{\text{A}} \over \mu_{\text{B}}} > {\nu_{\text{A}} \over \nu_{\text{B}}}$ ("case 2"). Allowed regions are marked in color for either cases. In case 1, we show that ${{\partial Y_{11}^a} \over {\partial \mu_{\text{B}}}} <0$, so any given point A can descend along $\mu_{\text{B}}$ axis (the solid black arrow) to get higher rate, until it reaches boundary line ${\mu_{\text{A}} \over \mu_{\text{B}}} = {\nu_{\text{A}} \over \nu_{\text{B}}}$ where $\mu_{\text{B}}$ is highest. Similarly, in case 2, ${{\partial Y_{11}^b} \over {\partial \mu_{\text{A}}}} <0$, so any given point B can descend along $\mu_{\text{A}}$ axis until ${\mu_{\text{A}} \over \mu_{\text{B}}} = {\nu_{\text{A}} \over \nu_{\text{B}}}$ to get highest rate. Therefore, the optimal $(\mu_{\text{A}}^{opt}, \mu_{\text{B}}^{opt})$ that maximize the piecewise function $Y_{11}^{X,L}$ always occur on the boundary line. Moreover, for any given point $C(\mu_{\text{A}},\mu_{\text{B}})$ on the boundary line, the function values of $Y_{11}^a$, $Y_{11}^b$ are the same. However, we show that ${{\partial Y_{11}^a} \over {\partial \mu_{\text{A}}}}$ at $C_1$ is not equal to ${{\partial Y_{11}^b} \over {\partial \mu_{\text{A}}}}$ at $C_2$ (along the dot-dash red lines). Therefore, the piecewise function $Y_{11}^{X,L}$ is not smooth.}
	\label{fig:Y11}
\end{figure*}

To prove Theorem I, here we will actually propose an even stronger assumption for $\mu_{\text{A}}, \mu_{\text{B}}$:\\

\textit{\textbf{Theorem III}.
	for any arbitrary choice of device and channel parameters, and any two given values of $\nu_{\text{A}}, \nu_{\text{B}}$, the optimal decoy intensities $\mu_{\text{A}}^{opt}, \mu_{\text{B}}^{opt}$ that maximize $R$ always satisfy the constraint:}

\begin{equation}
{\mu_{\text{A}}^{opt} \over \mu_{\text{B}}^{opt} } = { \nu_{\text{A}} \over \nu_{\text{B}}}
\end{equation}

\textit{\textbf{Remark:}} as will be shown below, Theorem I is simply a corollary from Theorem III.\\

\textit{\textbf{Proof for Theorem III:}} Here for convenience, we first limit the discussion to asymptotic case (i.e. infinite data size), and we assume that the vacuum intensity is indeed $\omega=0$. Throughout the rest of the text, we will use $Q_{ij}^k$ and $E_{ij}^k$ to denote the observed gain and QBER, where, if not specified, the first subscript is Alice's intensity, and the second is Bob's intensity, which can be chosen from $\{s_{\text{A}},\mu_{\text{A}},\nu_{\text{A}},\omega\}$ and $\{s_{\text{B}},\mu_{\text{B}},\nu_{\text{B}},\omega\}$ for Alice and Bob, respectively. The superscript $k$ signifies the basis X or Z (although, here we only explicitly write the basis for illustration purposes, since the basis is already implied by the choice of intensities).

First, looking at the key rate expression \cite{mdiqkd,mdifourintensity}:

\begin{equation}
\begin{aligned}
R=P_{s_{\text{A}}}P_{s_{\text{B}}} \{(s_{\text{A}} e^{-s_{\text{A}}})(s_{\text{B}} e^{-s_{\text{B}}}) Y_{11}^{X,L}[1-h_2(e_{11}^{X,U})]\\
-f_eQ_{ss}^Z h_2(E_{ss}^Z)\}
\end{aligned}
\end{equation}

\noindent we can see that only the term $Y_{11}^{X,L}[1-h_2(e_{11}^{X,U})]$, i.e. the decoy-state analysis and privacy amplification, is determined by the decoy intensities (and probabilities, if finite-size effect is considered) only, and not affected by the signal intensities $s_{\text{A}}, s_{\text{B}}$. This is an important and very convenient characteristic of the 4-intensity/7-intensity protocol, that the signal state is only concerned with key generation, while the decoy states are only responsible for privacy amplification. That is, the optimization of decoy-state intensities is decoupled from the key generation. Now, we can make an observation that, under given device parameters and channel loss, the optimization of the decoy intensities is independent of $s_{\text{A}}, s_{\text{B}}$, and its only goal is to maximize $Y_{11}^{X,L}[1-h_2(e_{11}^{X,U})]$.

Furthermore, to perform the decoy state analysis, we note that the upper bound for single-photon QBER satisfies the form of:

\begin{equation}
\begin{aligned}
e_{11}^{X,U} = f(Y_{11}^{X,L}, \nu_{\text{A}}, \nu_{\text{B}})
\end{aligned}
\end{equation}

\noindent where $e_{11}^{X,U}$ is only determined by $Y_{11}^{X,L}$, $\nu_{\text{A}}$ and $\nu_{\text{B}}$. The full expression, as in \cite{mdipractical}, is listed below:

\begin{equation}
\begin{aligned}
e_{11}^{X,U} = {1\over {\nu_{\text{A}}\nu_{\text{B}} Y_{11}^{X,L}}}(e^{\nu_{\text{A}}+\nu_{\text{B}}}Q_{\nu\nu}^XE_{\nu\nu}^X-e^{\nu_{\text{A}}}Q_{\nu\omega}^XE_{\nu\omega}^X\\
-e^{\nu_{\text{B}}}Q_{\omega\nu}^XE_{\omega\nu}^X+Q_{\omega\omega}^XE_{\omega\omega}^X)
\end{aligned}
\end{equation}.

Now, suppose we first fix two arbitrary values of $\nu_{\text{A}},\nu_{\text{B}}$, and try to maximize $Y_{11}^X[1-h_2(e_{11}^X)]$ by optimizing $\mu_{\text{A}}, \mu_{\text{B}}$, we can see that maximizing $Y_{11}^{X,L}$ will suffice, since it will simultaneously minimize $e_{11}^{X,U}$, whose only component dependent on $\mu_{\text{A}}, \mu_{\text{B}}$ is $Y_{11}^{X,L}$. The question now becomes simply finding:

\begin{equation}
({\mu_{\text{A}}^{opt}, \mu_{\text{B}}^{opt} }) = argmax(Y_{11}^{X,L}(\mu_{\text{A}},\mu_{\text{B}}))
\end{equation}

A very important characteristic of $Y_{11}^{X,L}$ is that, its expression is dependent upon whether ${\mu_{\text{A}} \over \mu_{\text{B}}} \leq {\nu_{\text{A}} \over \nu_{\text{B}}}$, i.e. it is a piecewise function, as described in Ref.\cite{mdipractical}:\\

\textbf{Case 1}: If ${\mu_{\text{A}} \over \mu_{\text{B}}} \leq {\nu_{\text{A}} \over \nu_{\text{B}}}$:

\begin{equation}
Y_{11}^{X,L} = Y_{11}^a = {1 \over (\mu_{\text{A}}-\nu_{\text{A}})} [{{\mu_{\text{A}}}\over{\nu_{\text{A}}\nu_{\text{B}}}}Q_{\nu\nu}^{M1}-{{\nu_{\text{A}}}\over{\mu_{\text{A}}\mu_{\text{B}}}}Q_{\mu\mu}^{M2}]
\end{equation}

\textbf{Case 2}: otherwise, if ${\mu_{\text{A}} \over \mu_{\text{B}}} > {\nu_{\text{A}} \over \nu_{\text{B}}}$:

\begin{equation}
Y_{11}^{X,L} = Y_{11}^b = {1 \over (\mu_{\text{B}}-\nu_{\text{B}})} [{{\mu_{\text{B}}}\over{\nu_{\text{A}}\nu_{\text{B}}}}Q_{\nu\nu}^{M1}-{{\nu_{\text{B}}}\over{\mu_{\text{A}}\mu_{\text{B}}}}Q_{\mu\mu}^{M2}]
\end{equation}

\noindent where we denote the two expressions of $Y_{11}^{X,L}$ in the two cases as $Y_{11}^a$ and $Y_{11}^b$, and the two terms $Q_{\nu\nu}^{M1}$ and $Q_{\mu\mu}^{M2}$ are linear combinations of the observable Gain, and are functions of $(\nu_{\text{A}},\nu_{\text{B}})$ and $(\mu_{\text{A}},\mu_{\text{B}})$ only, respectively. Their full expressions can be found in Appendix F.3. Also, note that if ${\mu_{\text{A}} \over \mu_{\text{B}}} = {\nu_{\text{A}} \over \nu_{\text{B}}}$, the two cases $Y_{11}^a=Y_{11}^b$.

Now, we can make a key observation, that in case 1, for any given $\mu_{\text{A}}$, the partial derivative ${\partial Y_{11}^a} \over {\partial \mu_{\text{B}}}$ always satisfies

\begin{equation}
{{\partial Y_{11}^a} \over {\partial \mu_{\text{B}}}} < 0
\end{equation}

\noindent (The actual expression of the partial derivative and proof of its positivity are shown in Appendix F.3). However, in case 1, $\mu_{\text{B}}$ is bounded by $\mu_{\text{B}} \geq {{\mu_{\text{A}} \nu_{\text{B}}} \over \nu_{\text{A}}}$, so the only optimal case is to take the boundary condition

\begin{equation}
\mu_{\text{B}}^{opt} = {{\mu_{\text{A}} \nu_{\text{B}}} \over \nu_{\text{A}}}
\end{equation}

This means that, in the region of ${\mu_{\text{A}} \over \mu_{\text{B}}} \leq {\nu_{\text{A}} \over \nu_{\text{B}}}$, any two optimal value pair $(\mu_{\text{A}}^{opt},\mu_{\text{B}}^{opt})$ must satisfy ${\mu_{\text{A}}^{opt} \over \mu_{\text{B}}^{opt} } = { \nu_{\text{A}} \over \nu_{\text{B}}}$, or else we can always decrease $\mu_{\text{B}}$ to get a higher rate, meaning that the previous point is not the actual maximum. We illustrate this behavior in Fig.\ref{fig:Y11}.

Similarly, for case 2, the partial derivative with respect to $\mu_{\text{A}}$ satisfies

\begin{equation}
{{\partial Y_{11}^b} \over {\partial \mu_{\text{A}}}} < 0
\end{equation}

\noindent and $\mu_{\text{A}}$ is bounded by $\mu_{\text{A}} > {{\mu_{\text{B}} \nu_{\text{A}}} \over \nu_{\text{B}}}$. In the same way, in case 2 for any given $\mu_{\text{B}}$, we can acquire:

\begin{equation}
\mu_{\text{A}}^{opt} = {{\mu_{\text{B}} \nu_{\text{A}}} \over \nu_{\text{B}}}
\end{equation}

Up to here, we have proven that Theorem III is indeed correct.  \qedsymbol \\

\textit{\textbf{Proof for Theorem I:}} Now, following the same idea, any four optimal value pair $(\mu_{\text{A}}^{opt},\mu_{\text{B}}^{opt},\nu_{\text{A}}^{opt},\nu_{\text{B}}^{opt})$ must satisfy ${\mu_{\text{A}}^{opt} \over \mu_{\text{B}}^{opt} } = { \nu_{\text{A}}^{opt} \over \nu_{\text{B}}^{opt}}$, or else we can always vary $(\mu_{\text{A}}, \mu_{\text{B}})$ while keeping $(\nu_{\text{A}}, \nu_{\text{B}})$ fixed, and let ${\mu_{\text{A}} \over \mu_{\text{B}}} = {\nu_{\text{A}} \over \nu_{\text{B}}}$ to get a higher rate, meaning that the previous point is not the actual maximum. Therefore, we have shown that Theorem I is indeed correct, that the optimal decoy intensities always satisfy

\begin{equation}
{\mu_{\text{A}}^{opt} \over \mu_{\text{B}}^{opt} } = { \nu_{\text{A}}^{opt} \over \nu_{\text{B}}^{opt}}
\end{equation}

\qedsymbol\\

Note that the same conclusion doesn't hold true for traditional 3-intensity MDI-QKD (i.e. using $\{\mu, \nu, \omega\}$ for both X and Z basis and using $\mu$ in Z basis to generate the key), that is because the key rate for 3-intensity depends on $\mu$ for both key generation and error-correction, such that $Q_{\mu\mu}^Z, E_{\mu\mu}^Z$ terms and the single-photon probability $\mu e^{-\mu}$ both depend on $\mu$, hence optimizing only $Y_{11}^L$ is no longer sufficient. Therefore, this independence of $s$ from $\mu, \nu$ is an additional advantage that the 4-intensity/7-intensity protocol can provide, under asymmetric conditions.

Also, one thing to note is that, although the above theorem provides us with a way to constrain ${\mu_{\text{A}} \over \mu_{\text{B}}}, {\nu_{\text{A}} \over \nu_{\text{B}}}$, the actual values of these ratios still need to be found by optimization. In Ref. \cite{mdipractical}, the authors have proposed a rule-of-thumb formula for finding optimal intensities:

\begin{equation}
\mu_{\text{A}} \eta_{\text{A}} = \mu_{\text{B}} \eta_{\text{B}}
\end{equation}

\noindent for which we now have a good understanding of the reason - such a relation keeps the arriving intensities balanced at Charles, in order to maintain good HOM visibility in the X basis and low QBER.

However, this is still only a rough approximation, and is an exact relation only when the dark count rate $Y_0$ is ignored, data size is infinite, and infinite number of decoys are used (Ref. \cite{mdipractical} considered the case where $\mu$ is both the signal and decoy intensity, and only proved Eq. (E13) to be exact in the ideal infinite-decoy case with no noise). For a general case, $\mu_{\text{A}}/\mu_{\text{B}}$ is not always exactly equal to $\eta_{\text{B}}/\eta_{\text{A}}$ (and does not only depend on the mismatch $x=\eta_{\text{A}}/\eta_{\text{B}}$) but rather deviates slightly from it when $(\eta_{\text{A}},\eta_{\text{B}})$ changes. But at least, one general rule is that $\mu_{\text{A}}/\mu_{\text{B}}$ decreases with $x=\eta_{\text{A}}/\eta_{\text{B}}$, or, to put in more simple words, the larger the channel loss, the higher the intensities we should choose to compensate for the loss.

\subsection{Non-smoothness of Key Rate vs Intensities Function}

In the previous section we have shown that the piecewise expression for $Y_{11}^{X,L}$ causes the optimal value to occur on the boundary line ${\mu_{\text{A}} \over \mu_{\text{B}}} = {\nu_{\text{A}} \over \nu_{\text{B}}}$. Here we continue to show that Theorem II is a result of this piecewise function, too.\\

\textit{\textbf{Proof of Theorem II:}} The theorem means that, the key rate does not have a continuous partial derivative with respect to $\mu_{\text{A}}$ or $\mu_{\text{B}}$ at the boundary line. This will cause the boundary line to behave like a sharp "ridge". To prove this, instead of differentiating $Y_{11}^a$ vs $\mu_{\text{B}}$ and $Y_{11}^b$ vs $\mu_{\text{A}}$, here we perform partial differentiation of both $Y_{11}^a$, $Y_{11}^b$ vs $\mu_{\text{A}}$, and observe this discontinuity of derivative.

First, we rewrite $Y_{11}$ into:

\begin{equation}
\begin{aligned}
Y_{11}^a = {\nu_{\text{A}} \over (\mu_{\text{A}}-\nu_{\text{A}})} [{{1}\over{\nu_{\text{A}}\nu_{\text{B}}}}Q_{\nu\nu}^{M1}-{{1}\over{\mu_{\text{A}}\mu_{\text{B}}}}Q_{\mu\mu}^{M2}] + {{1}\over{\nu_{\text{A}}\nu_{\text{B}}}}Q_{\nu\nu}^{M1}\\
Y_{11}^b = {\nu_{\text{B}} \over (\mu_{\text{B}}-\nu_{\text{B}})} [{{1}\over{\nu_{\text{A}}\nu_{\text{B}}}}Q_{\nu\nu}^{M1}-{{1}\over{\mu_{\text{A}}\mu_{\text{B}}}}Q_{\mu\mu}^{M2}] + {{1}\over{\nu_{\text{A}}\nu_{\text{B}}}}Q_{\nu\nu}^{M1}\\
\end{aligned}
\end{equation}

The last term is not dependent on either $\mu_{\text{A}}$ or $\mu_{\text{B}}$. Note that, here on the boundary of ${\mu_{\text{A}} \over \mu_{\text{B}}} = {\nu_{\text{A}} \over \nu_{\text{B}}}$, the values of $Y_{11}^a$ and $Y_{11}^b$ are equal:

\begin{equation}
Y_{11}^a=Y_{11}^b
\end{equation}

Performing the partial differentiation against $\mu_{\text{A}}$, we can get:

\begin{equation}
\begin{aligned}
{{\partial Y_{11}^a} \over {\partial \mu_{\text{A}}}} =  &-{\nu_{\text{A}} \over{\mu_{\text{A}}-\nu_{\text{A}}}} {{\partial} \over {\partial \mu_{\text{A}}}}({{Q_{\mu\mu}^{M2}}\over {\mu_{\text{A}} \mu_{\text{B}}}}) \\
&+ {\nu_{\text{A}} \over{(\mu_{\text{A}}-\nu_{\text{A}})^2}} ({{Q_{\mu\mu}^{M2}}\over {\mu_{\text{A}} \mu_{\text{B}}}}-{{Q_{\nu\nu}^{M1}}\over {\nu_{\text{A}} \nu_{\text{B}}}})\\
{{\partial Y_{11}^b} \over {\partial \mu_{\text{A}}}} =  &-{\nu_{\text{B}} \over{\mu_{\text{B}}-\nu_{\text{B}}}} {{\partial} \over {\partial \mu_{\text{A}}}}({{Q_{\mu\mu}^{M2}}\over {\mu_{\text{A}} \mu_{\text{B}}}})\\
\end{aligned}
\end{equation}

We can see that, on the boundary of ${\mu_{\text{A}} \over \mu_{\text{B}}} = {\nu_{\text{A}} \over \nu_{\text{B}}}$, the first terms are again equal, however, the second term in ${{\partial Y_{11}^a} \over {\partial \mu_{\text{A}}}}$ is strictly larger than 0 (detailed proof by expanding $Q_{\mu\mu}^{M2}$, $Q_{\nu\nu}^{M1}$ are shown in Appendix F.3). Therefore,

\begin{equation}
{{\partial Y_{11}^a} \over {\partial \mu_{\text{A}}}} \neq {{\partial Y_{11}^b} \over {\partial \mu_{\text{A}}}}
\end{equation}

The derivatives of $Y_{11}^{X,L}$ vs $\mu_{\text{A}}$ on the two sides of the "ridge" are not equal, causing the rate function R to be have a non-defined gradient. A similar proof can be applied to $\mu_{\text{B}}$ and it leads to the same result. \qedsymbol\\

An illustration can be seen in main text Fig. 2, which chooses a given set of values $(\nu_{\text{A}}=0.2, \nu_{\text{B}}=0.1)$ and plots the key rate over $(\mu_{\text{A}},\mu_{\text{B}})$. As can be clearly observed, there is a sharp ridge on the line ${\mu_{\text{A}} \over \mu_{\text{B}}}={\nu_{\text{A}} \over \nu_{\text{B}}}=2 $, meaning the key rate function versus intensities is not smooth.

\subsection{Proof of Negativity of Partial Derivatives for Decoy Intensities}

As described above, the expression for the single-photon yield, $Y_{11}^{X,L}$ depends on whether ${\mu_{\text{A}} \over \mu_{\text{B}}} \leq {\nu_{\text{A}} \over \nu_{\text{B}}}$. For case 1, if ${\mu_{\text{A}} \over \mu_{\text{B}}} \leq {\nu_{\text{A}} \over \nu_{\text{B}}}$, we would like to prove that

\textit{\textbf{Lemma I:} ${\partial Y_{11}^a} \over {\partial \mu_{\text{B}}}$ and ${\partial Y_{11}^b} \over {\partial \mu_{\text{A}}}$ are both always negative.\\ }

\textit{\textbf{Proof of Lemma I:}} Here, we use a simplified model of the Gain $Q_{ij}^X$ as in Ref.\cite{mdipractical}, which ignores the dark count rate $Y_0$, and takes a second-order approximation for the modified Bessel function:

\begin{equation}
Q_{\mu\mu}^X={\eta_{\text{B}}^2 \eta_d^2 \over 4} [2x\mu_{\text{A}}\mu_{\text{B}} + (\mu_{\text{B}}^2+x^2\mu_{\text{A}}^2)(2e_d-e_d^2)]
\end{equation}

\noindent where $\eta_{\text{B}}$ is the transmittance in Bob-Charles channel, $x={\eta_{\text{A}} \over \eta_{\text{B}}}$ is the channel mismatch, $\eta_d$ is the detector efficiency, and $e_d$ is the misalignment. Here for convenience we can further define

\begin{equation}
\begin{aligned}
\epsilon &= 2e_d - e_d^2,  \,\,\,\,\,\,\,\,\,\,\,\, T = {\eta_{\text{B}}^2 \eta_d^2 \over 4}
\end{aligned}
\end{equation}

\noindent such that
\begin{equation}
Q_{\mu\mu}^X=T (2x\mu_{\text{A}}\mu_{\text{B}} + \epsilon\mu_{\text{B}}^2 +\epsilon x^2\mu_{\text{A}}^2)
\end{equation}

Now, let us consider  ${\partial Y_{11}^a} \over {\partial \mu_{\text{B}}}$ where

\begin{equation}
Y_{11}^{X,L} = Y_{11}^a = {1 \over (\mu_{\text{A}}-\nu_{\text{A}})} [{{\mu_{\text{A}}}\over{\nu_{\text{A}}\nu_{\text{B}}}}Q_{\nu\nu}^{M1}-{{\nu_{\text{A}}}\over{\mu_{\text{A}}\mu_{\text{B}}}}Q_{\mu\mu}^{M2}]
\end{equation}

To calculate the single-photon gain, the two terms:

\begin{equation}
\begin{aligned}
Q_{\nu\nu}^{M1} &= e^{\nu_{\text{A}}+\nu_{\text{B}}}Q_{\nu\nu}^X - e^{\nu_{\text{A}}}Q_{\nu\omega}^X - e^{\nu_{\text{B}}}Q_{\omega\nu}^X + Q_{\omega\omega}^X\\
Q_{\mu\mu}^{M2} &= e^{\mu_{\text{A}}+\mu_{\text{B}}}Q_{\mu\mu}^X - e^{\mu_{\text{A}}}Q_{\mu\omega}^X - e^{\mu_{\text{B}}}Q_{\omega\mu}^X + Q_{\omega\omega}^X\\
\end{aligned}
\end{equation}

\noindent are linear combinations of the observable Gains $Q_{ij}^Z$.

We can make the observation that, only the term
\begin{equation}
-{\nu_{\text{A}} \over {(\mu_{\text{A}}-\nu_{\text{A}})\mu_{\text{A}}}}\left({Q_{\mu\mu}^{M2}\over \mu_{\text{B}}}\right)
\end{equation}

\noindent contains $\mu_{\text{B}}$, so, we only need to prove the \textit{positivity} of ${\partial \over {\partial \mu_{\text{B}}}} ({Q_{\mu\mu}^{M2}\over \mu_{\text{B}}})$, where

\begin{equation}
\begin{aligned}
Q_{\mu\mu}^{M2} &= e^{\mu_{\text{A}}+\mu_{\text{B}}}Q_{\mu\mu}^X - e^{\mu_{\text{A}}}Q_{\mu\omega}^X - e^{\mu_{\text{B}}}Q_{\omega\mu}^X + Q_{\omega\omega}^X\\
&= e^{\mu_{\text{A}}+\mu_{\text{B}}}Q_{\mu\mu}^X - e^{\mu_{\text{A}}}Q_{\mu\omega}^X - e^{\mu_{\text{B}}}Q_{\omega\mu}^X\\
\end{aligned}
\end{equation}

\noindent substituting with Eq. (E20),

\begin{equation}
\begin{aligned}
{1\over T}Q_{\mu\mu}^{M2} &= (2x\mu_{\text{A}}\mu_{\text{B}}+x^2\epsilon \mu_{\text{A}}^2 + \epsilon \mu_{\text{B}}^2)e^{\mu_{\text{A}}+\mu_{\text{B}}} \\
&- x^2\epsilon \mu_{\text{A}}^2 e^{\mu_{\text{A}}} - \epsilon \mu_{\text{B}}^2 e^{\mu_{\text{B}}}\\
&= 2x\mu_{\text{A}}\mu_{\text{B}} e^{\mu_{\text{A}}+\mu_{\text{B}}} \\
&+ x^2\epsilon \mu_{\text{A}}^2 e^{\mu_{\text{A}}}(e^{\mu_{\text{B}}}-1) + \epsilon \mu_{\text{B}}^2 e^{\mu_{\text{B}}} (e^{\mu_{\text{A}}}-1)\\
\end{aligned}
\end{equation}

\noindent Therefore,

\begin{equation}
\begin{aligned}
{Q_{\mu\mu}^{M2} \over {\mu_{\text{B}}}} &= T \times [2x\mu_{\text{A}} e^{\mu_{\text{A}}+\mu_{\text{B}}} + x^2\epsilon \mu_{\text{A}}^2 e^{\mu_{\text{A}}}{{e^{\mu_{\text{B}}}-1}\over \mu_{\text{B}}} \\
&+ \epsilon(e^{\mu_{\text{A}}}-1) \mu_{\text{B}} e^{\mu_{\text{B}}}]\\
\end{aligned}
\end{equation}

\noindent note that here, as $\mu_{\text{A}}, \mu_{\text{B}}>0$, we have $e^{\mu_{\text{A}}}, e^{\mu_{\text{B}}} > 1$, and each of the three functions satisfy
\begin{equation}
\begin{aligned}
{\partial \over {\partial \mu_{\text{B}}}} (e^{\mu_{\text{A}}+\mu_{\text{B}}}) &>0 \\
{\partial \over {\partial \mu_{\text{B}}}} ({{e^{\mu_{\text{B}}}-1}\over \mu_{\text{B}}}) &>0 \\
{\partial \over {\partial \mu_{\text{B}}}} (\mu_{\text{B}} e^{\mu_{\text{B}}}) &>0 \\
\end{aligned}
\end{equation}

Therefore, we have proven that ${\partial \over {\partial \mu_{\text{B}}}} ({Q_{\mu\mu}^{M2}\over \mu_{\text{B}}}) >0$ and that ${{\partial Y_{11}^a} \over {\partial \mu_{\text{B}}}} <0$. Similarly, we can also prove that  ${{\partial Y_{11}^b} \over {\partial \mu_{\text{A}}}} <0$. Thus, the optimal point $(\mu_{\text{A}}^{opt},\mu_{\text{B}}^{opt})$ must happen on the boundary, i.e.
\begin{equation}
{\mu_{\text{A}}^{opt} \over \mu_{\text{B}}^{opt} } = { \nu_{\text{A}} \over \nu_{\text{B}}}
\end{equation}

\qedsymbol

\subsection{Proof of Discontinuity of Partial Derivatives for Decoy Intensities}

Now, to prove the discontinuity of the first-order derivatives of the key rate function, we need to show that\\

\textit{\textbf{Lemma II:} Partial derivative of $Y_{11}^a$ and $Y_{11}^b$ with respect to $\mu_{\text{A}}$, i.e. ${\partial Y_{11}^a} \over {\partial \mu_{\text{A}}}$ and ${\partial Y_{11}^b} \over {\partial \mu_{\text{A}}}$ are not equal.}

\textit{\textbf{Proof of Lemma II:}}

\begin{equation}
\begin{aligned}
{{\partial Y_{11}^a} \over {\partial \mu_{\text{A}}}} =  &-{\nu_{\text{A}} \over{\mu_{\text{A}}-\nu_{\text{A}}}} {{\partial} \over {\partial \mu_{\text{A}}}}({{Q_{\mu\mu}^{M2}}\over {\mu_{\text{A}} \mu_{\text{B}}}}) \\
&+ {\nu_{\text{A}} \over{(\mu_{\text{A}}-\nu_{\text{A}})^2}} ({{Q_{\mu\mu}^{M2}}\over {\mu_{\text{A}} \mu_{\text{B}}}}-{{Q_{\nu\nu}^{M1}}\over {\nu_{\text{A}} \nu_{\text{B}}}})\\
{{\partial Y_{11}^b} \over {\partial \mu_{\text{A}}}} =  &-{\nu_{\text{B}} \over{\mu_{\text{B}}-\nu_{\text{B}}}} {{\partial} \over {\partial \mu_{\text{A}}}}({{Q_{\mu\mu}^{M2}}\over {\mu_{\text{A}} \mu_{\text{B}}}})\\
\end{aligned}
\end{equation}

On the boundary of ${\mu_{\text{A}} \over \mu_{\text{B}}} = {\nu_{\text{A}} \over \nu_{\text{B}}}$, the first terms are equal, i.e.

\begin{equation}
-{\nu_{\text{A}} \over{\mu_{\text{A}}-\nu_{\text{A}}}} {{\partial} \over {\partial \mu_{\text{A}}}}({{Q_{\mu\mu}^{M2}}\over {\mu_{\text{A}} \mu_{\text{B}}}}) = -{\nu_{\text{B}} \over{\mu_{\text{B}}-\nu_{\text{B}}}} {{\partial} \over {\partial \mu_{\text{A}}}}({{Q_{\mu\mu}^{M2}}\over {\mu_{\text{A}} \mu_{\text{B}}}})
\end{equation}

\noindent However, here we have to show that the second term in ${{\partial Y_{11}^a} \over {\partial \mu_{\text{A}}}}$ is strictly larger than zero:

\begin{equation}
{\nu_{\text{A}} \over{(\mu_{\text{A}}-\nu_{\text{A}})^2}} ({{Q_{\mu\mu}^{M2}}\over {\mu_{\text{A}} \mu_{\text{B}}}}-{{Q_{\nu\nu}^{M1}}\over {\nu_{\text{A}} \nu_{\text{B}}}}) > 0
\end{equation}

\noindent or, since $\mu_{\text{A}} > \nu_{\text{A}}$ and $\mu_{\text{A}}, \nu_{\text{A}} >0$, simply

\begin{equation}
{{Q_{\mu\mu}^{M2}}\over {\mu_{\text{A}} \mu_{\text{B}}}}-{{Q_{\nu\nu}^{M1}}\over {\nu_{\text{A}} \nu_{\text{B}}}} > 0
\end{equation}

Just like in Appendix F.3, we can expand the gain $Q_{ij}^X$ using Eq. (E20):

\begin{equation}
\begin{aligned}
{{Q_{\nu\nu}^{M1}}\over {\nu_{\text{A}} \nu_{\text{B}}}} =  2x e^{\nu_{\text{A}}+\nu_{\text{B}}} &+ x^2\epsilon {\nu_{\text{A}}\over \nu_{\text{B}}} e^{\nu_{\text{A}}}(e^{\nu_{\text{B}}}-1) \\
&+ \epsilon {\nu_{\text{B}} \over \nu_{\text{A}}} e^{\nu_{\text{B}}} (e^{\nu_{\text{A}}}-1)\\
{{Q_{\mu\mu}^{M2}}\over {\mu_{\text{A}} \mu_{\text{B}}}} =  2x e^{\mu_{\text{A}}+\mu_{\text{B}}} &+ x^2\epsilon {\mu_{\text{A}}\over \mu_{\text{B}}} e^{\mu_{\text{A}}}(e^{\mu_{\text{B}}}-1) \\
&+ \epsilon {\mu_{\text{B}} \over \mu_{\text{A}}} e^{\mu_{\text{B}}} (e^{\mu_{\text{A}}}-1)\\
\end{aligned}
\end{equation}

\noindent Subtracting them, we can acquire:
\begin{equation}
\begin{aligned}
{{Q_{\mu\mu}^{M2}}\over {\mu_{\text{A}} \mu_{\text{B}}}}-{{Q_{\nu\nu}^{M1}}\over {\nu_{\text{A}} \nu_{\text{B}}}} &=  2x (e^{\mu_{\text{A}}+\mu_{\text{B}}}-e^{\nu_{\text{A}}+\nu_{\text{B}}}) \\
&+ x^2\epsilon (\mu_{\text{A}} e^{\mu_{\text{A}}} {{e^{\mu_{\text{B}}}-1}\over \mu_{\text{B}}} - \nu_{\text{A}} e^{\nu_{\text{A}}} {{e^{\nu_{\text{B}}}-1}\over \nu_{\text{B}}})\\
&+ \epsilon (\mu_{\text{B}} e^{\mu_{\text{B}}} {{e^{\mu_{\text{A}}}-1}\over \mu_{\text{A}}} - \nu_{\text{B}} e^{\nu_{\text{B}}} {{e^{\nu_{\text{A}}}-1}\over \nu_{\text{A}}})\\
\end{aligned}
\end{equation}

\noindent Note that, when a given variable $x>0$, the functions

\begin{equation}
\begin{aligned}
{d \over {d x}} (e^x) &>0 \\
{d \over {d x}} ({{e^{x}-1}\over x}) &>0 \\
{d \over {d x}} (x e^{x}) &>0 \\
\end{aligned}
\end{equation}

\noindent Therefore, these three functions strictly increase with their variable $x$, i.e. for any $x_1>x_2$, $f(x_1)>f(x_2)$. Now, we can use the conditions $\mu_{\text{A}} > \nu_{\text{A}}$, $\mu_{\text{B}} > \nu_{\text{B}}$, and acquire:

\begin{equation}
\begin{aligned}
e^{\mu_{\text{A}}+\mu_{\text{B}}} &> e^{\nu_{\text{A}}+\nu_{\text{B}}}\\
\mu_{\text{A}} e^{\mu_{\text{A}}} &> \nu_{\text{A}} e^{\nu_{\text{A}}}\\
\mu_{\text{B}} e^{\mu_{\text{B}}} &> \nu_{\text{B}} e^{\nu_{\text{B}}}\\
{{e^{\mu_{\text{A}}}-1}\over \mu_{\text{A}}} &> {{e^{\nu_{\text{A}}}-1}\over \nu_{\text{A}}}\\
{{e^{\mu_{\text{B}}}-1}\over \mu_{\text{B}}} &> {{e^{\nu_{\text{B}}}-1}\over \nu_{\text{B}}}\\
\end{aligned}
\end{equation}

Therefore, we have proven that ${{Q_{\mu\mu}^{M2}}\over {\mu_{\text{A}} \mu_{\text{B}}}}-{{Q_{\nu\nu}^{M1}}\over {\nu_{\text{A}} \nu_{\text{B}}}} > 0$, i.e.

\begin{equation}
{{\partial Y_{11}^a} \over {\partial \mu_{\text{A}}}} \neq {{\partial Y_{11}^b} \over {\partial \mu_{\text{A}}}}
\end{equation}

Similarly, one can show that

\begin{equation}
{{\partial Y_{11}^b} \over {\partial \mu_{\text{B}}}} \neq {{\partial Y_{11}^a} \over {\partial \mu_{\text{B}}}}
\end{equation}

Therefore, for any given intensities $(\nu_{\text{A}},\nu_{\text{B}})$, the rate function $R(\mu_{\text{A}},\mu_{\text{B}})$ is not smooth against the two intensities $(\mu_{\text{A}},\mu_{\text{B}})$. \qedsymbol\\

\textit{\textbf{Remark:}} Also, though not explicitly proven here - since $\nu_{\text{A}}, \nu_{\text{B}}$ will affect not only $Y_{11}^{X,L}$, but will affect $e_{11}^{X,U}$ too, their derivatives will be a lot more complex than $\mu_{\text{A}}, \mu_{\text{B}}$ - numerically we observed that for any given $(\mu_{\text{A}},\mu_{\text{B}})$, the rate function $R(\nu_{\text{A}},\nu_{\text{B}})$ is actually not smooth against the two intensities $(\nu_{\text{A}},\nu_{\text{B}})$ either, and the ridge still appears at ${\mu_{\text{A}} \over \mu_{\text{B}}} = {\nu_{\text{A}} \over \nu_{\text{B}}}$.

\section{Local Search Algorithm}

In this section we describe how to perform the optimization for the parameters, which is an indispensable process in obtaining the optimal key rate. In addition, we also discuss the effect of inaccuracies and fluctuations of the intensities and probabilities on the key rate, and show that our method is robust even in the presence of inaccuracies and fluctuations of the parameters.

To provide a good key rate under finite-size effects, optimal choice of parameters is an extremely important factor in implementing the protocol. However, the 7-intensity protocol has an extremely large parameter space of 12 dimensions, for which a brute-force search is next to impossible. To put into context, a desktop PC (quad-core i7-4790k@4.0GHz) can evaluate the function $R(\vec{v})$ at approximately $10^5$ parameter combinations $\vec{v}$ per second. But searching over a very crude 10-sample resolution for each parameter would take over 4 months, and a 100-sample resolution for each parameter would take $3 \times 10^{11}$ years, a time longer than the age of the universe! Therefore, a local search algorithm must be used to efficiently search the parameters in reasonable time.

{\color{black} There have been studies to apply convex optimization to QKD e.g. in Ref. \cite{mdiparameter} to find the optimal set of parameters and in Refs. \cite{convex1,convex2,convex3} to bound the information leakage and secure key rate.} Here we start by adopting a local search algorithm for parameter optimization, proposed in Ref. \cite{mdiparameter}, called "coordinate descent" (CD), which requires drastically less time than using an exhaustive search. Instead of performing an exhaustive search over the parameter space, we can descend along each axis at a time, and iterate over each axis in turn. For instance, suppose we currently iterate $s_{\text{A}}$:

\begin{equation}
\begin{aligned}
R^{i+1}=\max_{s_{\text{A}} \in ({s_{\text{A}}}^{min},{s_{\text{A}}}^{max})} &R(s_{\text{A}}, \mu_{\text{A}}^i, \nu_{\text{A}}^i, P_{s_{\text{A}}}^i, P_{\mu_{\text{A}}}^i, P_{\nu_{\text{A}}}^i,\\
&s_{\text{B}}^i, \mu_{\text{B}}^i, \nu_{\text{B}}^i, P_{s_{\text{B}}}^i, P_{\mu_{\text{B}}}^i, P_{\nu_{\text{B}}}^i) \\
= &R(s_{\text{A}}^{i+1}, \mu_{\text{A}}^i, \nu_{\text{A}}^i, P_{s_{\text{A}}}^i, P_{\mu_{\text{A}}}^i, P_{\nu_{\text{A}}}^i, \\
&s_{\text{B}}^i, \mu_{\text{B}}^i, \nu_{\text{B}}^i, P_{s_{\text{B}}}^i, P_{\mu_{\text{B}}}^i, P_{\nu_{\text{B}}}^i)
\end{aligned}
\end{equation}

\noindent which freezes the other coordinates, and replaces $s_{\text{A}}$ with the optimal position on the current coordinate-axis $s_{\text{A}}$. In the next iteration the algorithm will descent along axis $\mu_{\text{A}}$, etc., hence the name coordinate descent. The search space satisfies that: the probabilities lie within $(0,1)$, and while the intensities could be in principle larger than 1, typically that doesn't provide a good key rate, so here we also define the domain for all intensities as $(0,1)$. The decoy intensities also follow two additional constraints $\mu_{\text{A}} > \nu_{\text{A}}$ and $\mu_{\text{B}} > \nu_{\text{B}}$. The CD algorithm is able to reach the same optimal position as a gradient descent algorithm (with descends along the gradient vector), the commonly used approach for parameter optimization.

However, a significant limitation of coordinate descent is that it does not work correctly over functions that have discontinuous first-order derivatives (which cause the gradient to be non-defined). For instance, in the presence of a sharp "ridge" as in Fig.2 in the main text, any arbitrary point $P$ on the ridge will cause the CD algorithm to terminate incorrectly and fail to find the maximum point. Mathematically, this is caused by the gradient being not clearly defined at a position where derivatives are discontinuous. Therefore, coordinate descent does not work anymore for asymmetric MDI-QKD.

As we discussed above, such discontinuity of derivatives comes from the "ridge", ${\mu_{\text{A}} \over \mu_{\text{B}}}={\nu_{\text{A}} \over \nu_{\text{B}}}$. Moreover, we know that the optimal parameters must satisfy ${\mu_{\text{A}}^{opt} \over \mu_{\text{B}}^{opt} } = { \nu_{\text{A}}^{opt} \over \nu_{\text{B}}^{opt}}$. Therefore, here we propose to use polar coordinate instead of Cartesian coordinate to perform coordinate descent, and \textit{jointly} search ${\mu_{\text{A}} \over \mu_{\text{B}}}$ and ${\nu_{\text{A}} \over \nu_{\text{B}}}$. In this way, we can make the rate vs parameter function smooth. We redefine $\vec{v}$ as:

\begin{equation}
\vec{v_{polar}}=[s_{\text{A}}, s_{\text{B}}, r_\mu, r_\nu, \theta_{\mu\nu}, P_{s_{\text{A}}}, P_{\mu_{\text{A}}},P_{\nu_{\text{A}}},P_{s_{\text{B}}},P_{\mu_{\text{B}}},P_{\nu_{\text{B}}}]
\end{equation}

\noindent where
\begin{equation}
\begin{aligned}
r_\mu&=\sqrt{\mu_{\text{A}}^2+\mu_{\text{B}}^2}  \,\,\,\,\,\,\,\,\,\,\,\,		r_\nu=\sqrt{\nu_{\text{A}}^2+\nu_{\text{B}}^2} \\
\theta_{\mu\nu}&=tan^{-1}(\mu_{\text{A}}/\mu_{\text{B}})=tan^{-1}(\nu_{\text{A}}/\nu_{\text{B}})\\
\end{aligned}
\end{equation}

\noindent In this way, the expression of $Y_{11}^{L}$ always takes the boundary value (and only has a single expression). Therefore, when other parameters are fixed, $R(\theta_{\mu\nu})$ is actually a smooth function, therefore by searching over the parameters $\vec{v_{polar}}$, we can successfully find the optimal parameters and maximum rate.

After converting to polar coordinates and jointly searching $\theta_{\mu\nu}$, the coordinate descent algorithms becomes:

\begin{equation}
\begin{aligned}
R^{i+1}=\max_{s_{\text{A}} \in ({s_{\text{A}}}^{min},{s_{\text{A}}}^{max})} &R(s_{\text{A}}, s_{\text{B}}^i, r_\mu^i, r_\nu^i, \theta_{\mu\nu}^i,\\
& P_{s_{\text{A}}}^i, P_{\mu_{\text{A}}}^i,P_{\nu_{\text{A}}}^i,P_{s_{\text{B}}}^i,P_{\mu_{\text{B}}}^i,P_{\nu_{\text{B}}}^i) \\
= &R(s_{\text{A}}^{i+1}, s_{\text{B}}^i, r_\mu^i, r_\nu^i, \theta_{\mu\nu}^i,\\
& P_{s_{\text{A}}}^i, P_{\mu_{\text{A}}}^i,P_{\nu_{\text{A}}}^i,P_{s_{\text{B}}}^i,P_{\mu_{\text{B}}}^i,P_{\nu_{\text{B}}}^i) \\
\end{aligned}
\end{equation}

{\color{black}
Additionally, when searching along each coordinate (for instance, fixing other parameters and searching $s_{\text{A}}$), we employ an iterative searching technique to further accelerate the algorithm, which starts out with a coarse resolution and iteratively narrows the search region while increasing the resolution (this is a similar technique as introduced in Ref. \cite{mdiparameter}, but efficiently parallelized to utilize multi-threading on modern PCs). For instance, we can start out with e.g. 100 samples within the $(0,1)$ region and evaluate them in parallel. After the maximal point is found, we can then choose two neighboring samples on the left and right of the maximal point, and start a finer search among 10 more samples between them. This process can be iterated until maximum value no longer changes significantly, or until maximum depth is reached. Such technique allows a search resolution that dynamically changes as needed (from $10^{-2}$ down to even $10^{-5}$, although in practice often $10^{-3}$ is sufficient), and it efficiently uses e.g. the 8 threads on a quad-core CPU, enabling fast and accurate optimization below 0.1s.

One more note is that, the key rate obtained by our method is in fact robust against small inaccuracies in the parameters. For instance, for Point A3 (10km, 60km) in Table III, if we round all parameters to an accuracy of 0.001 (as shown in Table IV) and use it for simulation, we can still get $99.5\%$ of the optimal key rate $3.106\times 10^{-5}$, while rounding the parameters to 0.01 will still give us $93.0\%$ of the optimal key rate. In fact, even if we just keep one significant digit of each parameter, we can still get $47.6\%$ of the optimal key rate. This would make it much easier for an experimental implementation of our method, as the key rate is very forgiving of inaccuracies in the parameters, which makes a much less stringent requirement on the intensity modulators and random number generators.

Note that, the above "accuracy" discusses how strict the requirement is for us to generate an intensity/probability with mean value close to the desired optimal value (e.g. limited by bits in the random number generator or the accuracy of the intensity modulator), but we are still assuming we have perfect knowledge of the variables we generate. In addition, here we would like to point out that our conclusions remain unchanged, even in the presence of intensity fluctuations, or imprecision in the intensity probabilities. 

Firstly, the system is not very sensitive to the probabilities (since the partial derivatives with respect to them are zero at the optimal points), so even if all signal and decoy probabilities are simultaneously set 5\% away from optimal value (and we take the global worst-case key rate value among all possible combinations of positive/negative deviation for each variable), the key rate will not significantly drop - for instance for the (10km, 60km) case, one can still obtain 92.3\% the ideal key rate ($2.869\times 10^{-5}$ versus $3.106\times 10^{-5}$) even with a 5\% deviation for the probabilities. 

Similarly, for intensity fluctuations, even if we add a 5\% deviation to \textit{all} intensities (again, taking the (10km, 60km) case as an example) we can still get 73.1\% the ideal key rate ($2.270\times 10^{-5}$ versus $3.106\times 10^{-5}$). Moreover, one important point to note is that, intensity fluctuation is not a problem unique to asymmetric MDI-QKD (or the new asymmetric protocol that we propose in this work). Even if one uses prior protocols (such as the 4-intensity protocol), one would still obtain a significantly lower key rate if taking intensity fluctuation into consideration, such as 39.9\% the key rate ($3.671\times 10^{-5}$ versus $9.206\times 10^{-6}$ with no fluctuation) at (0km, 50km), and zero key rate (versus $3.891\times 10^{-7}$ with no fluctuation) at (10km, 60km). Therefore, the advantage of our method remains unchanged, even if intensity fluctuations are considered.

}
\section{Finite Size analysis}

In this section we describe the finite-key analysis used in our simulations.

The analytical proofs in Appendix F are shown for the asymptotic case. Numerically we show that 7-intensity protocol works effectively in the finite-key regime too, as can be observed in main text Fig. \ref{fig:2d_Results}.

To account for finite-size effects, we perform a standard error analysis\cite{mdifourintensity,mdiparameter}, and estimate the expected value $\langle n \rangle$ of an observable $n$ by

\begin{equation}
	\begin{aligned}
		\underline{n} = n - \gamma \sqrt{n} \leq \langle n \rangle \leq  n + \gamma \sqrt{n} = \overline{n}
	\end{aligned}
\end{equation} 

\noindent where we define the upper and lower bound for an observable $n$ as $\overline{n}$ and $\underline{n}$. Here, $\gamma$ is the number of standard deviations the confidence interval of the observed value is from the expected value (for instance, for a required failure probability of no more than $\epsilon=10^{-7}$, we should set $\gamma=5.3$).

We can denote the observed counts as $n_{\mu_i,\mu_j}^X$, and error counts as $m_{\mu_i,\mu_j}^X$, where $\mu_i \in \{\mu_{\text{A}}, \nu_{\text{A}}, \omega\}$, $\mu_j \in \{\mu_{\text{B}}, \nu_{\text{B}}, \omega\}$. Then, the observed gain and error can be acquired from:

\begin{equation}
	\begin{aligned}
		Q_{\mu_i,\mu_j}^X &= {{n_{\mu_i,\mu_j}^X} \over {NP_{\mu_i}P_{\mu_j}}} \\
		T_{\mu_i,\mu_j}^X &= {{m_{\mu_i,\mu_j}^X} \over {NP_{\mu_i}P_{\mu_j}}} \\
		E_{\mu_i,\mu_j}^X &= {{T_{\mu_i,\mu_j}^X} \over {Q_{\mu_i,\mu_j}^X}} \\
	\end{aligned}
\end{equation}

\noindent where $N$ is the total number of signals sent, and $P_{\mu_i},P_{\mu_j}$ are the probabilities for Alice and Bob to send the respective intensities. Note that here we define the QBER in terms of error-gains: 

\begin{equation}
	T_{\mu_i,\mu_j}^X = Q_{\mu_i,\mu_j}^X E_{\mu_i,\mu_j}^X
\end{equation}

As described in Appendix F, we can define the key rate expression as \cite{mdiqkd,mdifourintensity}:
\begin{equation}
	\begin{aligned}
		R=P_s P_s \{s_{\text{A}} s_{\text{B}} e^{-(s_{\text{A}}+s_{\text{B}})}Y_{11}^{X,L}[1-h_2(e_{11}^{X,U})]\\
		-f_e Q_{ss}^Z h_2(E_{ss}^Z)\}
	\end{aligned}
\end{equation} 

\noindent and the single-photon gain and error estimated by \cite{mdipractical}:

\begin{equation}
	\begin{aligned}
		Y_{11}^{X,L}= {1\over {\mu_{\text{A}}-\nu_{\text{A}}}} ({{\mu_{\text{A}}}\over {\nu_{\text{A}}\nu_{\text{B}}}} Q_{\nu\nu}^{M1} - {{\nu_{\text{A}}}\over {\mu_{\text{A}}\mu_{\text{B}}}} Q_{\mu\mu}^{M2})\\
		e_{11}^{X,U} = {1\over {\nu_{\text{A}}\nu_{\text{B}}Y_{11}^{X,L}}}(e^{\nu_{\text{A}}+\nu_{\text{B}}}T_{\nu\nu}-e^{\nu_{\text{A}}}T_{\nu\omega}\\
		-e^{\nu_{\text{B}}}T_{\omega\nu}+T_{\omega\omega})\\
	\end{aligned}
\end{equation}

\noindent where $Q_{\nu\nu}^{M1}, Q_{\mu\mu}^{M2}$ are linear combination terms of the observables

\begin{equation}
	\begin{aligned}
		Q_{\nu\nu}^{M1} = e^{\nu_{\text{A}}+\nu_{\text{B}}}Q_{\nu\nu}^X - e^{\nu_{\text{A}}}Q_{\nu\omega}^X - e^{\nu_{\text{B}}}Q_{\omega\nu}^X + Q_{\omega \omega}^X\\
		Q_{\mu\mu}^{M2} = e^{\mu_{\text{A}}+\mu_{\text{B}}}Q_{\mu\mu}^X - e^{\mu_{\text{A}}}Q_{\mu\omega}^X - e^{\mu_{\text{B}}}Q_{\omega\mu}^X + Q_{\omega \omega}^X\\
	\end{aligned}
\end{equation}

Now, with standard error analysis, we can define the upper and lower bounds for the gain and error-gain:

\begin{equation}
	\begin{aligned}
		\overline{Q_{\mu_i\mu_j}^X}=Q_{\mu_i\mu_j}^X + \gamma \sqrt{Q_{\mu_i\mu_j}^X \over {N P_{\mu_i} P_{\mu_j}}} \\
		\underline{Q_{\mu_i\mu_j}^X}=Q_{\mu_i\mu_j}^X - \gamma \sqrt{Q_{\mu_i\mu_j}^X \over {N P_{\mu_i} P_{\mu_j}}} \\
		\overline{T_{\mu_i\mu_j}^X}=T_{\mu_i\mu_j}^X + \gamma \sqrt{T_{\mu_i\mu_j}^X \over {N P_{\mu_i} P_{\mu_j}}} \\
		\underline{T_{\mu_i\mu_j}^X}=T_{\mu_i\mu_j}^X - \gamma \sqrt{T_{\mu_i\mu_j}^X \over {N P_{\mu_i} P_{\mu_j}}} \\
	\end{aligned}
\end{equation}

\noindent 

\noindent Therefore, we have

\begin{equation}
	\begin{aligned}
		\underline{Q_{\nu\nu}^{M1}} = e^{\nu_{\text{A}}+\nu_{\text{B}}}\underline{Q_{\nu\nu}^X} - e^{\nu_{\text{A}}}\overline{Q_{\nu\omega}^X} - e^{\nu_{\text{B}}}\overline{Q_{\omega\nu}^X} + \underline{Q_{\omega \omega}^X}\\
		\overline{Q_{\mu\mu}^{M2}} = e^{\mu_{\text{A}}+\mu_{\text{B}}}\overline{Q_{\mu\mu}^X} - e^{\mu_{\text{A}}}\underline{Q_{\mu\omega}^X} - e^{\mu_{\text{B}}}\underline{Q_{\omega\mu}^X} + \underline{Q_{\omega \omega}^X}\\
		Y_{11}^{X,L}= {1\over {\mu_{\text{A}}-\nu_{\text{A}}}} ({{\mu_{\text{A}}}\over {\nu_{\text{A}}\nu_{\text{B}}}} \underline{Q_{\nu\nu}^{M1}} - {{\nu_{\text{A}}}\over {\mu_{\text{A}}\mu_{\text{B}}}} \overline{Q_{\mu\mu}^{M2}}) \\
		e_{11}^{X,U} = {1\over {\nu_{\text{A}}\nu_{\text{B}}Y_{11}^{X,L}}}(e^{\nu_{\text{A}}+\nu_{\text{B}}}\overline{T_{\nu\nu}}-e^{\nu_{\text{A}}}\underline{T_{\nu\omega}}\\
		-e^{\nu_{\text{B}}}\underline{T_{\omega\nu}}+\overline{T_{\omega\omega}})\\
	\end{aligned}
\end{equation}

\noindent which we can use to substitute into Eq. (G4) to obtain the key rate under finite-size effects. (Note that here $Q_{\omega \omega}^X$ takes the lower bound in both $\underline{Q_{\nu\nu}^{M1}}$ and $\overline{Q_{\mu\mu}^{M2}}$, because its overall coefficient is positive in $Y_{11}^{X,L}$).

\begin{table*}[t]
	\caption{{\color{black}Simulation results of key rate estimated with independent-bounds versus joint-bounds, using parameters in Table I. The data points for independent-bounds correspond to the solid red curve in Fig.\ref{fig:2d_Results} (d). As can be seen, using joint-bounds for finite-size estimation can improve the key rate significantly. However, this will result in multiple maxima and cause instabilities in simulations. Therefore, we have used independent-bounds throughout the main text.}}
	\begin{center}
		{\color{black}
		\begin{tabular}{cccc}			
			\hline \hline
			$L_{\text{A}}$ & $L_{\text{B}}$ & $R_{\text{independent}}$ & $R_{\text{joint}}$\\
			\hline
			$60km$ & $10km$ & $3.106\times 10^{-5}$ & $6.714\times 10^{-5}$ \\
			$100km$ & $50km$ & $4.677\times 10^{-11}$ & $7.568\times 10^{-8}$ \\
			$113km$ & $63km$ & $0$ & $7.311\times 10^{-10}$ \\
			\hline \hline
			
		\end{tabular}
		}
	\end{center}
\end{table*}

\begin{figure}[h]
	\includegraphics[scale=0.4]{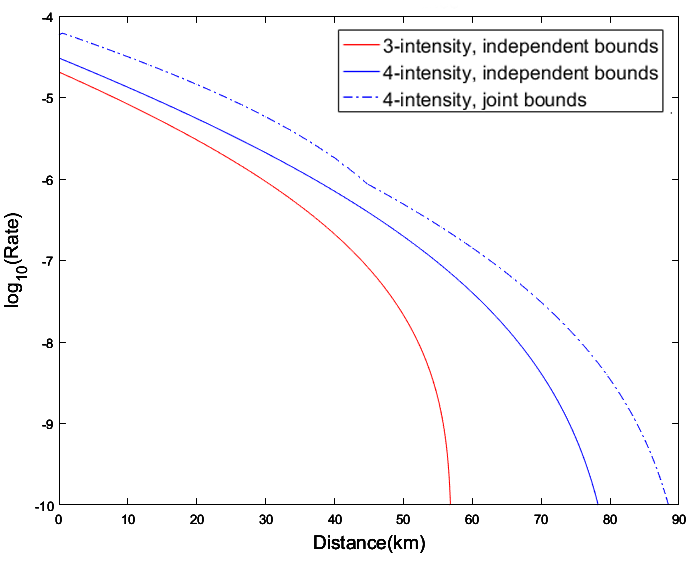}
	\caption{Rate vs distance (Alice to Bob) for symmetric case, for $N=10^{11}$ using parameters $Y_0=6.02\times 10^{-6},\eta_d=14.5\%,e_d=1.5\%$, a parameter set in Zhou et al.'s paper \cite{mdifourintensity}. Here we compare the traditional 3-intensity protocol as proposed in Ref.\cite{mdiparameter} (red solid line), and the 4-intensity protocol\cite{mdifourintensity} with independent-bound (blue solid line) and joint-bound analysis (blue dot-dash line).}
	\label{fig:rateN11}
\end{figure}

Note that, in Ref. \cite{mdifourintensity}, in addition to proposing the 4-intensity protocol, Zhou et al. has proposed a "joint-bounds" finite-key analysis which jointly considers the statistical fluctuations of observable Gain and QBER. It is a tighter bound and can provide higher rate than considering each observable's fluctuation independently as we've discussed above in this section (i.e. using "independent-bounds"). To illustrate this, we perform a simple simulation of key rate versus distance plot, using independent-bounds and joint-bounds (as well as using traditional 3-intensity protocol \cite{mdiparameter} for comparison). As can be seen in Fig.\ref{fig:rateN11}, 4-intensity protocol with joint-bounds analysis provides higher rate than independent-bounds (and both have higher rate than the 3-intensity protocol). 

However, joint-bound analysis is based on linear optimization and sometimes brings multiple maxima for $R(\vec{v})$, which is undesirable for local search, and will result in unpredictable behaviors (such as sudden "jitters" in the resulting rate versus distance plot, as can be observed in the joint-bound plot in Fig.\ref{fig:rateN11}. Similar behavior is observed in Ref.\cite{mdifourintensity} too). 

{\color{black}
Here just for comparison, we list in Table V some example data points where we apply both independent-bound and joint-bound analysis. As can be seen, using joint-bounds, we can indeed gain a further improved key rate. However, this comes at the expense of not knowing whether we are indeed at the global maximum or not, due to the existence of multiple maxima (and is not ideal for comparing asymmetric/symmetric protocols, as the key rate estimated could be just local maxima for both of them). Therefore, as the purpose of this work is studying asymmetric MDI-QKD, we focus on independent-bounds throughout the main text. 
}

Also, note that although we have used standard error-analysis for simplicity, our method here can in principle be applied to finite-key analysis with composable security, too, such as using Chernoff bound \cite{mdiChernoff}. The key point is that (as explicitly demonstrated in Appendices B and C), the scaling of asymmetric MDI-QKD key rate versus distances depends on the signal states (which performs a trade-off between error-correction and single photon probability). The decoy states need to maintain balanced arriving intensities at Charles, but only serve to estimate the single-photon contributions as accurately as possible, whose asymptotic bounds are given by the infinite-data, infinite-decoy case. Adopting different finite-key analysis (or no analysis at all, as in asymptotic case) affects the bounds on single photon gain and QBER $Y_{11}^L$ and $e_{11}^U$. The finite-size case can be seen as the asymptotic case with correction terms (i.e. imperfections) added to the privacy amplification, but its key rate will have a similar scaling property as the asymptotic case. This means that the advantage of our method is independent of the finite-size analysis model used (or lack thereof, in the asymptotic case).

\section{Single-Arm MDI-QKD}
\begin{table*}[t]
	\caption{Simulation results of key rate between each pair of nodes in a MDI-QKD network, using parameters from Table I, $N=10^{11}$, and channels in main text Fig. 1(a). As can be seen, using 7-intensity protocol always provides higher rate than either using 4-intensity directly (which fails to establish some connections) or using 4-intensity after adding fibre to each channel to accommodate the longest channel (which results in identical low rate for every connection - since every channel equals the longest channel after adding fibre). 7-intensity protocol therefore enables high scalability and reconfigurability because each link is independent of other links and no added fibre is needed.}
	\begin{center}
		\begin{tabular}{ccccccc}			
			\hline \hline
			Method & $A_1$-$A_3$ & $A_1$-$A_4$ & $A_1$-$A_5$ & $A_3$-$A_4$ & $A_3$-$A_5$ & $A_4$-$A_5$\\
			\hline
			4-intensity, add fibre & $1.28\times 10^{-10}$& $1.28\times 10^{-10}$& $1.28\times 10^{-10}$& $1.28\times 10^{-10}$& $1.28\times 10^{-10}$& $1.28\times 10^{-10}$\\
			4-intensity, direct & 0 & 0 & 0 & $2.41\times 10^{-4}$& $3.22\times 10^{-4}$& $5.77\times 10^{-4}$\\
			7-intensity, direct &$1.97\times 10^{-7}$& $2.42\times 10^{-7}$& $2.77\times 10^{-7}$ & $2.48\times 10^{-4}$& $3.53\times 10^{-4}$& $5.87\times 10^{-4}$\\
			\hline \hline
			
		\end{tabular}
	\end{center}
\end{table*}

\begin{figure}[h]
	\includegraphics[scale=0.3]{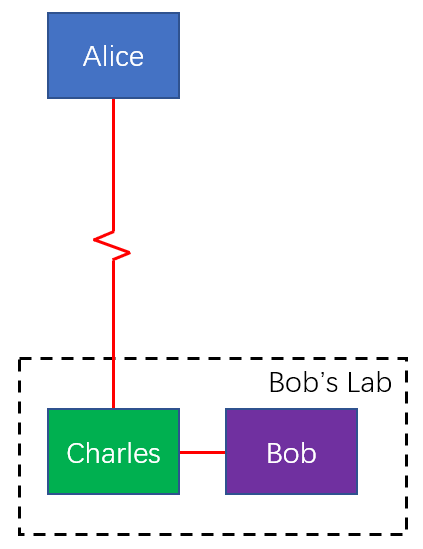}
	\caption{"Single-arm" MDI-QKD where Bob and Charles are both in the same lab, with Bob's channel having as little loss as possible. By optimizing intensities, we can achieve maximum distance (loss) in the single channel between Alice and Charles, while enjoying the security of MDI-QKD.}
	\label{fig:zero_arm}
\end{figure}

\begin{figure}[h]
	\includegraphics[scale=0.4]{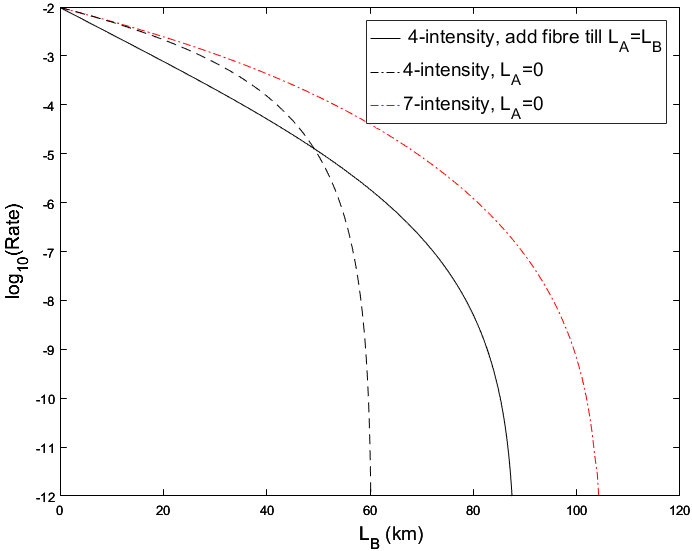}
	\caption{Simulations of "single-arm" MDI-QKD. We use parameters from Table I, and set $N=10^{11}$. The three lines are generated using 4-intensity protocol and adding fibre until $L_{\text{A}}=L_{\text{B}}$ (black solid line), using 4-intensity  protocol but without being able to add fibre (black dashed line), and using 7-intensity protocol directly (red dot-dash line). As can be seen, using 7-intensity protocol tremendously increases the key rate and maximum distance for the longer single-arm. At $R=10^{-7}$, using 7-intensity protocol (having maximum distance at 90km) increases maximum distance by 17.5 or 33.2km (or, 3.5 to 6.6dB of loss) compared to 4-intensity with/without fibre, respectively.}
	\label{fig:singlearm}
\end{figure}

In the main text we have proposed a new type of "single-arm" MDI-QKD setup, which is the extremely asymmetric case where one channel has high loss while the other channel has close to zero loss. In this section we will describe it in more detail and outline its potential applications.

Suppose we have one crucial channel (e.g. a free-space channel, say in a satellite-ground connection, or a ship-to-ship connection) through which we would like to send quantum signals. We would like to prevent all attacks on the detector and improve the security with MDI-QKD, but cannot add a third party in the middle of the free-space channel. In this case, it is possible to add another source Bob in the laboratory (alongside Charles' detectors, with as small loss as possible in Bob-Charles channel), and use it to interfere with the signals coming from Alice over the longer free-space channel, as shown in Fig.\ref{fig:zero_arm}. With 7-intensity protocol, high key rate can be generated from this extremely asymmetric case, providing the security of MDI-QKD to a single channel where relays cannot be added while still maintaining good performance.

If one uses 4-intensity protocol, Bob has to add a fibre similar in loss to that of the free-space channel (to maintain the symmetry), while as we've shown with 7-intensity protocol, Bob can simply choose as small a loss as possible, and obtain maximum acceptable loss in Alice's channel. Not only does 7-intensity protocol make such a highly asymmetric MDI-QKD possible, it actually provides a higher rate compared to the symmetric case (if Bob adds a fibre). Moreover, since Alice's channel loss might be constantly changing, it can be very difficult to adjust an added fibre and maintain the symmetry, thus the convenience of not having to add any loss with 7-intensity protocol is a significant factor, too.

As we can observe in main text Fig. \ref{fig:2d_Results}(a)(b), for the same required minimum rate, rather than performing an experiment at $(L_{max},L_{max})$, if we are free to adjust one channel (and want maximum distance in the other channel), we can set the shorter channel to zero, and obtain a longer distance in the other channel, e.g. $(L_{max}',0)$ with $L_{max}'>L_{max}$. For instance, in main text Fig. \ref{fig:2d_Results}(a)(b), choosing point $B(102km,0km)$ can extend the longer arm from 85km to 102km, from the symmetric point $A(85km,85km)$ for the same $R=10^{-10}$.

Here we list the simulations results for single-arm MDI-QKD. To demonstrate the advantage, here we study three-cases: using 4-intensity (but being able to add fibre until channels are symmetric), using 4-intensity (however, due to being e.g. in a free-space channel or a dynamic network, without the luxury to add fibres and compensate for the channels), and using 7-intensity directly on the asymmetric channels. As can be seen in Fig.\ref{fig:singlearm}, 7-intensity protocol provides better performance that both strategies using 4-intensity, and increases maximum distance from 56.8km and 72.5km (respectively for adding/not adding fibre) to 90km. Thus, our new protocol can enable a unique new application of providing the security of MDI-QKD to a single channel where relays cannot be added (e.g. a free-space link), while still maintaining high key rate.\\

\section{MDI-QKD Network Numerical Results}

In this section we consider the channels from a real quantum network setup in Vienna, reported in Ref.\cite{quantumnetwork1}, and numerically show that using 7-intensity protocol can provide high-rate communication between each pair of users, while previous protocols either fail to establish some connections in the network, or suffer from low key rate for all connections.

Here, we focus here on the high-asymmetry nodes in Ref.\cite{quantumnetwork1}, $A_1,A_2,A_3,A_4,A_5$, plotted in main text Fig. 1(a), and consider the case where an untrusted relay is placed at $A_2$. The topology here is a commonly studied model of a star-type network, which is considered for QKD network in \cite{starnetwork1,starnetwork2}, and is also the model for the MDI-QKD network experiment in Ref.\cite{mdinetwork}. Such a network can provide a complete graph of connections between any two users, but only requires one physical connection from each user. We show the simulation results in Table VI, where using 7-intensity protocol consistently provides high-rate connections even for nodes with very high asymmetry, and maintains the same (in fact moderately higher) key rate for nodes that are near-symmetric, i.e. including a long channel doesn't affect the rate between pairs of existing shorter channels.

Being able to establish connections with arbitrarily placed new nodes without affecting existing nodes is a very important property for a protocol to be used in a scalable and reconfigurable network, whose links will obviously be, more often than not, asymmetric. For the 4-intensity protocol, to accommodate the highest-loss channel, all connections will suffer from non-optimal key rate. Moreover, since new users might be added/deleted dynamically, such adding-fibre strategy will have poor scalability, since each new node affects the performance of all existing nodes, and also causes interruption of service when users update their fibres. With 7-intensity protocol, we are completely free of the worries of asymmetry, and can directly use the protocol on any channel combination optimally, so each node can be added/deleted without affecting the rest. This greatly improves not only the key rate, but also the scalability of a MDI-QKD network.

\end{document}